\documentstyle[12pt,graphicx,amsmath]{article}
\newlength{\pubnumber} 
\catcode`\@=11

\@addtoreset{equation}{section}
\def\section{\@startsection{section}{1}{\z@}{3.5ex plus 1ex minus .2ex}
 {2.3ex plus .2ex}{\large\bf}}
\def\subsection{\@startsection{subsection}{2}{\z@}{2.3ex plus .2ex}
 {2.3ex plus .2ex}{\bf}}
\numberwithin{equation}{section}

\RequirePackage{tabularx}
\RequirePackage{amssymb}

\newcommand{\B}[1]{B_{pqrs}^{(#1)}} 
\newcommand{\PP}[1]{P_{pqrs}^{(#1)}} 
\makeatletter
\renewcommand{\maketag@@@}[1]{\hbox{\m@th\normalsize\normalfont#1}}%
\makeatother

\begin{document}

\begin{titlepage}
\samepage{
\setcounter{page}{1}
\rightline{LTH--999}
\rightline{March 2014}

\vfill
\begin{center}
{\Large \bf{Classification\\ \medskip of Flipped $SU(5)$ Heterotic-String Vacua}}

\vspace{1cm}
\vfill 

{\large Alon E. Faraggi$^{1}$\footnote{E-mail address: alon.faraggi@liv.ac.uk} John Rizos$^{2}$\footnote{E-mail address: irizos@uoi.gr} and
Hasan Sonmez$^{1}\footnote{E-mail address: Hasan.Sonmez@liv.ac.uk}$}\\

\vspace{1cm}

{\it $^{1}$ Dept.\ of Mathematical Sciences, University of Liverpool, Liverpool L69 7ZL, UK\\}

\vspace{.05in}

{\it $^{2}$ Department of Physics, University of Ioannina, GR45110 Ioannina, Greece\\}

\vspace{.025in}
\end{center}

\vfill
\begin{abstract}
\noindent
We extend the classification of the free fermionic heterotic--string 
vacua to models in which the $SO(10)$ GUT symmetry at the string scale is broken to the flipped $SU(5)$ subgroup. 
In  our classification method, the set of basis vectors defined by the boundary 
conditions which are assigned to the free fermions is fixed and the enumeration of the
string vacua is obtained in terms of the
Generalised GSO (GGSO) projection coefficients entering the 
one--loop partition function. We derive algebraic expressions 
for the GGSO projections for all the physical states appearing in the
sectors generated by the set of basis vectors. 
This enables the analysis of the entire string spectrum to be programmed in to a computer code therefore, we performed a
statistical sampling in the space of $2^{44}\approx 10^{13}$
flipped $SU(5)$ vacua and scanned up to $10^{12}$ GGSO 
configurations. For that purpose, two independent 
codes were developed based on JAVA and FORTRAN95. All the results presented here are 
confirmed by the two independent routines. Contrary to the corresponding Pati--Salam
classification, we do not find exophobic flipped $SU(5)$
vacua with an odd number of generations. 
We study the structure of exotic states appearing in the
three generation models that additionally contain a viable
Higgs spectrum. Moreover, we demonstrate the existence of models 
in which all the exotic states are confined by a hidden
sector non--Abelian gauge symmetry as well as models 
that may admit the racetrack mechanism. 
\end{abstract}

\smallskip}
\end{titlepage}

\setcounter{footnote}{0}

\newcommand{\cc}[2]{c{#1\atopwithdelims[]#2}}
\newcommand{\nn}{\nonumber}

\def\AEF{A.E. Faraggi}
\def\MODA#1#2#3{{\it Mod.\ Phys.\ Lett.}\/ {\bf A#1} (#2) #3}
\def\IJMP#1#2#3{{\it Int.\ J.\ Mod.\ Phys.}\/ {\bf A#1} (#2) #3}
\def\nuvc#1#2#3{{\it Nuovo Cimento}\/ {\bf #1A} (#2) #3}
\def\RPP#1#2#3{{\it Rept.\ Prog.\ Phys.}\/ {\bf #1} (#2) #3}
\def\EJP#1#2#3{{\it Eur.\ Phys.\ Jour.}\/ {\bf C#1} (#2) #3}
\def\APP#1#2#3{{\it Astropart.\ Phys.}\/ {\bf #1} (#2) #3}
\def\JHEP#1#2#3{{\it JHEP}\/ {\bf #1} (#2) #3}
\def\NPB#1#2#3{{\it Nucl.\ Phys.}\/ {\bf B#1} (#2) #3}
\def\PLB#1#2#3{{\it Phys.\ Lett.}\/ {\bf B#1} (#2) #3}
\def\PRD#1#2#3{{\it Phys.\ Rev.}\/ {\bf D#1} (#2) #3}
\def\PRL#1#2#3{{\it Phys.\ Rev.\ Lett.}\/ {\bf #1} (#2) #3}
\def\PRT#1#2#3{{\it Phys.\ Rep.}\/ {\bf#1} (#2) #3}
\def\etal{{\it et al\/}}

\font\cmss=cmss10 \font\cmsss=cmss10 at 7pt

\hyphenation{space-time-super-sym-met-ric}
\hyphenation{su-per-sym-met-ric non-su-per-sym-met-ric}
\hyphenation{mod-u-lar mod-u-lar--in-var-i-ant}

\setcounter{footnote}{0}
\section{Introduction}
The LHC discovery of a Higgs--like resonance \cite{LHC2012}
lends further support to the viability of the Standard Model
as the effective parameterisation 
of all observational subatomic data up to the GUT or Planck scales.
This hypothesis is further supported by the proton lifetime,  
the neutrino mass suppression and the logarithmic evolution of the 
Standard Model parameters in the gauge and heavy generation matter sectors. 
The logarithmic evolution in the scalar sector is spoiled by
radiative corrections from the cutoff scale. Restoration of the
logarithmic running in the scalar sector suggests the 
existence of a new symmetry, with supersymmetry being
a concrete example of contemporary interest.

Despite its enormous success in accounting for observational 
subatomic data, the Standard Model is unsatisfactory.
It contains too many ad hoc parameters.
The gauge symmetries and representations 
are not selected by a fundamental principle.
The Standard Model\footnote{Including massive neutrinos.}
requires at least twenty--six additional parameters to account for the 
available data. 
The Standard Model gauge and flavour 
parameters can 
only be determined in a theory that unifies the 
gauge interactions with gravity. 
String theory provides a framework to study how the elementary particle's
attributes may arise from a consistent theory of gauge--gravity unification.
This necessitates the construction of quasi--realistic string models
and the investigation of their phenomenological properties. 
The tools assembled for this purpose
include target--space and worldsheet constructions \cite{IbanezUranga}.

The free fermionic formulation \cite{ABKKLT1987} of the heterotic--string in four
space--time dimensions provides a worldsheet approach to study 
quasi--realistic string vacua. The models constructed
in this formulation represent some of the most realistic 
string models corresponding
to symmetric and asymmetric ${\mathbb{Z}}_2 \times {\mathbb{Z}}_2$ orbifold compactifications.
Early examples of quasi--realistic free fermionic models which 
correspond to asymmetric  ${\mathbb{Z}}_2 \times {\mathbb{Z}}_2$ orbifold compactifications 
were constructed since the late eighties. The models
correspond to compactifications with ${\cal N}=(2,0)$
worldsheet supersymmetry in which the observable
$E_8$ symmetry is broken to a subgroup of $SO(10)$. 
The cases with
$SU(5)\times U(1)$ (flipped $SU(5)$) \cite{RevampAEHN,FSU51993}, 
$SO(6)\times SO(4)$ {Pati--Salam) \cite{ALR1990}, 
$SU(3)\times SU(2)\times U(1)^2$ (Standard--like) \cite{SLM} 
and 
$SU(3)\times SU(2)^2\times U(1)$ (left--right symmetric) \cite{LRS2001}
was shown to give rise to quasi--realistic examples.  

The early quasi--realistic free fermionic models consisted of a few
examples that shared an underlying NAHE--based structure \cite{NAHE1993}.
Contemporary research in string model building focuses on explorations
of large classes of string vacua. 
Over the last decade, tools for the classification of the symmetric ${\mathbb{Z}}_2 \times {\mathbb{Z}}_2$
free fermionic orbifolds were 
derived in \cite{GKR1999} for type II superstring 
and extended in \cite{FKNR2004, FKR2007}. 
Classification of the heterotic--string vacua with unbroken $SO(10)$ 
and $E_6$ GUT groups revealed the existence of
a symmetry in the space of ${\mathbb{Z}}_2$ and ${\mathbb{Z}}_2 \times {\mathbb{Z}}_2$ string models
under the exchange of spinorial plus anti--spinorial and vectorial
representations of $SO(10)$ \cite{FKR2007, CFKR2009}, which resembles
mirror symmetry \cite{Mirror1990}. The classification was extended 
to the string vacua in which the $SO(10)$ symmetry is broken to the 
Pati--Salam subgroup in \cite{ACFKR2011}.  It revealed 
the existence of exophobic string vacua, in which
fractionally charged states (exotics) appear in the massive spectrum, 
but do not exist among the massless states. A concrete three
generation exophobic model was studied in \cite{CFR2011}
and was shown to accommodate qualitatively viable phenomenology. 
The classification method in \cite{SU6SU2}, was used 
to fish out an exophobic model in which the $E_6$ symmetry is broken 
to the maximal $SU(6)\times SU(2)$ subgroup, which admits 
an additional family universal and anomaly free $U(1)$ symmetry
beyond the $U(1)$ generators of the $SO(10)$ GUT group
\cite{AnomalyFreeU1}.

The classification methodology developed in
\cite{GKR1999, FKNR2004, FKR2007, ACFKR2011},
provides a useful tool to explore the properties of 
large classes of string vacua. In this paper, we extend 
the classification to models in which the $SO(10)$ symmetry 
is broken to the flipped $SU(5)$ subgroup \cite{FSU5GUT1982}.
The novel aspect in the classification of this class of string models, 
is that the basis vectors that generate 
these contain boundary conditions that give rise to complex phases, whilst the 
models in the previous studies contained only periodic 
and antiperiodic boundary conditions. Extension of
the classification method to the flipped $SU(5)$ case 
is also a necessary step towards the classification of
Standard--like string vacua, which utilises both the
Pati--Salam and flipped $SU(5)$ generating basis vectors.
A question of particular interest in the classification is
the existence of quasi--realistic exophobic three generation
flipped $SU(5)$ models. We find that such a model does not
exist in the space of the order of $10^{12}$ that we explore. 
Our scan shows that the exophobic flipped $SU(5)$ models
exist in the string vacua with an even number of generations, but not in those with an odd number.

\section{Flipped $SU(5)$ Free Fermionic  Models}\label{analysis}
The quasi--realistic free fermionic models correspond to 
${\mathbb{Z}}_2 \times {\mathbb{Z}}_2$ orbifold compactifications 
with ${\cal N}=(2,0)$ super--conformal worldsheet symmetry. 
The free fermionic formulation \cite{ABKKLT1987} is set at an extended
symmetry point in the moduli space, where the
compactified directions are represented in terms of two
dimensional fermions propagating on the string
worldsheet \cite{Z2Z2Faraggi2002, Z2Z2Kounnas1997}.
Exactly marginal deformations from the
free fermionic point are obtained
by incorporating worldsheet Thirring
interactions among the worldsheet fermions \cite{Thirring1987}.
The free fermionic formulation provides a set of rules that
enables straightforward derivation of the physical states and
interactions plus is suited to explore the phenomenological
properties of the string vacua.
The matter states in the free fermionic models 
arise from the spinorial $\bf{16}$ representations of $SO(10)$, whilst the
Higgs states arise from the vectorial $\bf{1}0$ representation.
The ${\mathbb{Z}}_2 \times {\mathbb{Z}}_2$ free fermionic orbifold models therefore
preserve the Standard Model spectrum embedded in the $SO(10)$ GUT group.
The $SO(10)$ symmetry is broken at the string scale, leading to the gauge symmetry in the low energy effective
field theory being a subgroup of the $SO(10)$. 

In this paper, we extend the
classification  method to the case of the flipped $SU(5)$ subgroup. 
The distinctive feature of these models is the 
utilisation of rational boundary conditions, whereas the $SO(10)$ 
and $SO(6)\times SO(4)$ that were classified previously only used periodic and anti--periodic boundary conditions. Our classification method, entails that the GGSO projections for all
the states that arise in the twisted sectors
are expressed in terms of algebraic equations. The equations are
incorporated in a computer program that facilitates scanning a large 
number of models.

\subsection{Free Fermionic Formulation}
In the light-cone gauge, the four dimensional heterotic--string
is given by $44$ right--moving and $20$ left--moving real 
worldsheet fermions in the free fermionic construction. The models are constructed
by specifying the phases picked up by the
fermions $f_{1},\dots,f_{20},\overline{f}_{1},\dots,\overline{f}_{44}$
when parallel transported
along the non-contractible loops of the vacuum to vacuum amplitude.
The worldsheet fermions in the light-cone gauge in the usual
notation are:
$\psi^\mu, \chi^{1,\dots,6},y^{1,\dots,6}, \omega^{1,\dots,6}$
for the left-movers and
$\overline{y}^{1,\dots,6},\overline{\omega}^{1,\dots,6}$,
$\overline{\psi}^{1,\dots,5}$, $\overline{\eta}^{1,2,3}$, $\overline{\phi}^{1,\dots,8}$
for the right-movers. For a particular choice of
fermion phases, each model is defined to be consistent with the modular
invariance constraints, such that it can be spanned by a set of basis vectors
$v_{1},\dots,v_{N}$ given by

\begin{equation*}
v_i=\left\{\alpha_i(f_1),\dots,\alpha_i(f_{20})|\alpha_i(\overline{f}_1),
\dots,\alpha_i({\overline{f}_{44}})\right\}
\end{equation*}
describing the transformation properties of each worldsheet 
fermion

\begin{equation}
f_j\to -e^{i\pi\alpha_i(f_j)}\ f_j,  \,\,\,\,\,\,\,\,\,\, j = 1,\dots,64.\nonumber
\end{equation}
The basis vectors generate a space $\Xi$ that consists of
$2^{N+1}$ sectors which produce the string spectrum. Each
sector is given by a linear combination of the basis vectors

\begin{equation}
\xi = \sum_{i=1}^N m_j v_i, \,\,\,\,\,\,\,\,\,\, m_j = 0,1,\dots,N_j-1,\nonumber
\end{equation}
where $N_j \cdot v_j = 0$ mod $2$. 
The basis vectors induce the GGSO projections,
with action on a given string state $|S_{\xi}\rangle$ given by

\begin{equation}\label{gso}
e^{i\pi v_i\cdot F_{\xi}} |S_{\xi}\rangle = \delta_{{\xi}}\ C \binom {\xi} {v_i}^* |S_{\xi}\rangle,
\end{equation}
where $\delta_{{\xi}}=\pm1$
is the space--time spin statistics index
and $F_{\xi}$ is the fermion number operator.
Different sets of GGSO projection 
coefficients $\cc{{\xi}}{v_i}=\pm1; \pm i$,
consistent with modular invariance produce different models.
To summarize, a model is defined by a set of basis vectors
$v_{1},\dots,v_{N}$ and a set of $2^{N(N-1)/2}$ independent
GGSO projection coefficients $C \binom{v_i}{v_j}, \, i>j$ defining the string spectrum.

\subsection{SO(10) Models}
The flipped $SU(5)$ models we consider here are
generated by a set of 13 basis vectors. The first 12, consist of the same basis vectors
that were used in the classification of the $SO(10)$ vacua \cite{FKR2007}, which are given by
\begin{eqnarray}
v_1={\bf1}&=&\{\psi^\mu,\
\chi^{1,\dots,6},y^{1,\dots,6}, \omega^{1,\dots,6}| \nonumber\\
& & ~~~\overline{y}^{1,\dots,6},\overline{\omega}^{1,\dots,6},
\overline{\eta}^{1,2,3},
\overline{\psi}^{1,\dots,5},\overline{\phi}^{1,\dots,8}\},\nonumber\\
v_2=S&=&\{{\psi^\mu},\chi^{1,\dots,6}\},\nonumber\\
v_{2+i}={e_i}&=&\{y^{i},\omega^{i}|\overline{y}^i,\overline{\omega}^i\}, \
i=1,\dots,6,\nonumber\\
v_{9}={b_1}&=&\{\chi^{34},\chi^{56},y^{34},y^{56}|\overline{y}^{34},
\overline{y}^{56},\overline{\eta}^1,\overline{\psi}^{1,\dots,5}\},\label{basis}\\
v_{10}={b_2}&=&\{\chi^{12},\chi^{56},y^{12},y^{56}|\overline{y}^{12},
\overline{y}^{56},\overline{\eta}^2,\overline{\psi}^{1,\dots,5}\},\nonumber\\
v_{11}=z_1&=&\{\overline{\phi}^{1,\dots,4}\},\nonumber\\
v_{12}=z_2&=&\{\overline{\phi}^{5,\dots,8}\}.
\nonumber
\end{eqnarray}

The basis vectors ${\bf1}$ and $S$, generate a model
with the $SO(44)$ gauge symmetry and ${N} = 4$ 
space--time supersymmetry. The vectors $e_{1}$,$\dots$,$e_{6}$, corresponding
to all the possible symmetric shifts of the six internal
coordinates, break the $SO(44)$ gauge group to $SO(32) \times U(1)^6$ and preserve the ${N = 4}$
space--time supersymmetry.
The vectors $b_1$ and $b_2$ correspond to the ${\mathbb{Z}}_2 \times {\mathbb{Z}}_2$ orbifold twists, 
which break ${N} = 4$ to ${N} = 1$ supersymmetry
and reduces the rank of the group as the $U(1)^6$ is broken, leaving the $SO(32)$ symmetry to decompose to the $SO(10) \times U(1)^3 \times SO(16)$ gauge group. Furthermore, the $SO(10) \times U(1)^3$ group corresponds to our observable and the $SO(16)$ group to our hidden gauge group.
The remaining fermions that are not affected
by the action of the previous vectors are $\overline{\phi}^{1,\dots,8}$, 
which correspond to the $SO(16)$ gauge group. The vectors $z_1$ and $z_2$
reduce the 
untwisted hidden gauge group from $SO(16)$ to $SO(8)\times SO(8)$.
This choice of basis is the most general set of basis vectors
with symmetric shifts for the internal fermions
compatible with a $SO(10)$ GUT group. The untwisted vector bosons
consistent with the GGSO projections induced by the
choice of basis vectors in (\ref{basis}), generate the 
adjoint representation of an $SO(10)\times U(1)^3 \times SO(8)^2$
gauge group.

\subsection{Flipped $SU(5)$ Construction}
The $SO(10)$ GUT models generated by (\ref{basis}) are broken 
to the flipped $SU(5)$ subgroup, by the rational boundary condition assignment of the complex right-moving
fermions $\overline{\psi}^{1,\dots,5} = \pm \textstyle\frac{1}{2}$. This is achieved with the addition of the basis vector $v_{13} = \alpha$.
The case of the $SO(6)\times SO(4)$ models, which were classified
in \cite{ACFKR2011}, utilize solely periodic and anti--periodic
boundary conditions. In this case, the choice of $\alpha$ 
compatible with the set (\ref{basis}) is unique
and is given by $\alpha=\{{\overline\psi}^{4,5},{\overline\phi}^{1,2}\}$. All other 
possible assignments that reduce the $SO(10)$ symmetry to the 
$SO(6)\times SO(4)$ are equivalent. As in the cases of other
free fermionic flipped $SU(5)$ models constructed to date 
\cite{RevampAEHN,FSU51993}, we restrict the assignment of $\overline{\psi}^{1,\dots,5} $ to the case of positive
${1}/{2}$ boundary conditions. 
Furthermore, unlike the case of the $SO(6)\times SO(4)$ models, 
the choice of the basis vector $\alpha$ that breaks the 
$SO(10)$ symmetry to $SU(5)\times U(1)$ is not unique. 
Also, the assignment of the three complex worldsheet fermions
${\overline\eta}^{1,2,3}={1}/{2}$ is fixed by the modular invariance
constraint $b_j\cdot\alpha=0 \, {\rm mod} \, 1$. Consequently, it follows that the 
assignment of the boundary conditions of the eight worldsheet
complex fermions  ${\overline\psi}^{1,...,5}, {\overline\eta}^{1,2,3}$ 
is unique and the variation is in the boundary conditions of the 
worldsheet fermions ${\overline\phi}^{1,...,8}$. Modular invariance 
constraints, restrict the possibilities to assigning ${1}/{2}$
boundary conditions of ${\overline\phi}^{1,...,8}$ worldsheet fermions to 0, 4 or 8.
The null case been given by
\begin{equation}
\alpha=\{\overline{\psi}^{1,\dots,5}=
\textstyle\frac{1}{2},\,\overline{\eta}^{1,2,3}=\textstyle\frac{1}{2},\,
\overline{\phi}^{1,2}=1,\,\overline{\phi}^{3,4} = 1,\, 
\overline{\phi}^{5}=0,\,\overline{\phi}^{6,7}=0,\,\overline{\phi}^{8}=0\}.~~~~~\nonumber
\end{equation}
is automatically excluded because the sector
$x = 2\alpha = \{ \overline{\psi}^{1,\dots,5} , \overline{\eta}^{1,2,3} \}$
enhances the $SU(5)\times U(1)$ gauge group back to the $SO(10)$ symmetry.
The condition \\ $z_{1,2}\cdot\alpha=0 \, {\rm mod} \, 1$, imposes
the assignment of ${1}/{2}$ boundary conditions to 0, 2 or 4 
of each of the groups of worldsheet fermions
${\overline\phi}^{1,...,4}$ and ${\overline\phi}^{5,...,8}$. 
The possible choices of $v_{13}$ are then given by
\begin{eqnarray}
\alpha_1&=&\{\overline{\psi}^{1,\dots,5}=\textstyle\frac{1}{2},\,
\overline{\eta}^{1,2,3}=\textstyle\frac{1}{2},\,
\overline{\phi}^{1,2} = \textstyle\frac{1}{2},\,
\overline{\phi}^{3,4} = \textstyle\frac{1}{2},\,
\overline{\phi}^{5}=1,\,\overline{\phi}^{6,7}=0,\,
\overline{\phi}^{8}=0\,\},\nonumber\\
\alpha_2&=&\{\overline{\psi}^{1,\dots,5}=
\textstyle\frac{1}{2},\,\overline{\eta}^{1,2,3}=\textstyle\frac{1}{2},\,
\overline{\phi}^{1,2}=\textstyle\frac{1}{2},\,
\overline{\phi}^{3,4} = \textstyle\frac{1}{2},\,
\overline{\phi}^{5}=\textstyle\frac{1}{2},\, 
\overline{\phi}^{6,7}=\textstyle\frac{1}{2},\, 
\overline{\phi}^{8}=\textstyle\frac{1}{2} \}, \label{alphas}\\
\alpha_3&=&\{\overline{\psi}^{1,\dots,5}=
\textstyle\frac{1}{2},\,\overline{\eta}^{1,2,3}=\textstyle\frac{1}{2},\,
\overline{\phi}^{1,2}= \textstyle\frac{1}{2},\,
\overline{\phi}^{3,4} = 0,\, \overline{\phi}^{5}=1,\,
\overline{\phi}^{6,7}\,=\textstyle\frac{1}{2},\,\overline{\phi}^{8}=0\}. 
~~~~~
\nonumber
\end{eqnarray}
The $\alpha$'s above require that the sets of basis vectors are linearly
independent. This does not hold for the cases with $\alpha_1$ and $\alpha_2$,
since in these cases we obtain
\begin{eqnarray}
1 &=& S + \sum_{i = 0}^{6} e_i + 2 \alpha_1 + z_2 ,\nonumber\\
1 &=& S + \sum_{i = 0}^{6} e_i + 2 \alpha_2 + z_1 + z_2~.\nonumber 
\end{eqnarray}
In order to keep the set of basis vectors in (\ref{basis}), in addition to $\alpha_1$ or $\alpha_2$ being
linearly independent, we choose to remove the basis vector $\bf{1}$ leaving
the set of 12 vectors $\{ S,e_1,e_2,e_3,e_4,e_5,e_6,b_1,b_2,z_1,z_2,\alpha_i \}$,
where i = 1 or 2. In the case with $\alpha_3$, the set in
(\ref{basis}) is linearly independent giving us the set
$\{ 1,S,e_1,e_2,e_3,e_4,e_5,e_6,b_1,b_2,z_1,z_2,\alpha_3 \}$.
\\
\indent The plan for the remainder of the paper, is to give a comprehensive view   
of the methodology and an insight into the classification of the $SU(5) \times U(1)$ models with the inclusion of $\alpha_1$ in the 
basis. 
The classification
was carried out by using two independent codes, the first being the 
JAVA and the second being the FORTRAN95 code. It was then also carried out in the case
of $\alpha_2$. The details of the formulae needed for this classification
can be obtained from the authors and will be published
in a separate publication \cite{HasansThesis}. 
The classification using $\alpha_3$ will be reported in future work.

\subsection{GGSO projections}
In order to define the string vacua, the GGSO projection coefficients
appearing in the one-loop partition function $c\binom{v_i}{v_j}$ need
to be specified. Taking the coefficients to span a $12\times12$ matrix,
only the elements $i \geq j$ are independent. 
Modular invariance dictates that the 66 lower triangle elements of the
matrix are fixed by the corresponding 66 upper triangle elements.
Adding the remaining 12 diagonal terms, we are left with 78 independent
coefficients corresponding to $2^{78} \approx 3 \times 10^{23}$ different
string vacua. Moreover, requiring that the models possess ${N}=1$ space--time
supersymmetry, we fix eleven of the coefficients. Without loss of generality
we set the associated GGSO projection coefficients

\begin{eqnarray}
\label{Susyfixcoefficients}
C\binom{S}{S} = C\binom{S}{e_{i}} = C\binom{S}{b_{k}} =
C\binom{S}{z_1} = C\binom{S}{\alpha} = -1 , &&\\
i=1,...,6, \, k = 1,2.\nonumber
\end{eqnarray}
Modular invariance 
imposes  additional constraints on the diagonal terms.
In our case, where the vector $\bf{1}$ is composite, they are given by 

\begin{align}
C \binom {S} {z_2} &= - \prod_{\substack{i=1}}^{6}
C \binom {S} {e_i}, \nonumber\\
C \binom {e_k} {z_2} &= \; \; \; \, \prod_{\substack{i=1\\i \neq k}}^{6}
C \binom {e_k} {e_i}, \ \ \ k=1\ldots6, \nonumber\\
\label{constraint1}
C \binom {b_k} {b_k} &= - \prod_{i=1}^{6} C \binom {b_k} {e_i}
C \binom {b_k} {z_2}, \ \ \ k=1,2,\\
C \binom {z_1} {z_1} &= - \prod_{i=1}^{6} C \binom {z_1} {e_i}
C \binom {z_1} {z_2},\nonumber\\
C \binom {\alpha} {\alpha} &= - \prod_{i=1}^{6}
C \binom {\alpha} {e_i} C \binom {\alpha} {z_2},\nonumber
\end{align}
where $C \binom {z_2} {z_2}$ is independent of any term. Further analysis of the GGSO projections of interest, shows that
there are additional phases that do not
effect the properties of the string spectrum. As a result, 
the following coefficients are fixed in the ensuing analysis

\begin{eqnarray}
\label{freecoefficients}
C\binom{e_i}{e_i} = C\binom{e_3}{b_1} = C\binom{e_4}{b_1} =
C\binom{e_1}{b_2} = C\binom{e_2}{b_2} = C\binom{b_1}{b_2} =
C\binom{z_2}{z_2} = 1 , &&\\
\nonumber
\end{eqnarray}
where $i=1,..., 6$. Taking the equations 
(\ref{Susyfixcoefficients}), (\ref{constraint1}) and 
(\ref{freecoefficients}) we are left with 44 independent coefficients
which can take two discrete values $\pm1$, except in the cases
$C\binom{\alpha}{b_1}$, $C\binom{\alpha}{b_2}$ and
$C\binom{\alpha}{z_2}$, where they take the values $\pm i$ since
$\alpha\cdot b_1 = -3$ (odd),  $\alpha\cdot b_2 = -3$ (odd) and
$\alpha\cdot z_2 = - 1$ (odd). Furthermore, a simple counting
gives $2^{44} \approx 1.76 \times 10^{13}$
vacua in this
class of superstring models.
We note that there may still exist some degeneracies in this space
of vacua with regard to the characteristics of the low 
energy effective field theory, and in particular 
with respect to the observable massless states. 
For instance, 
the three twisted sectors of ${\mathbb{Z}}_2 \times {\mathbb{Z}}_2$ toroidal 
orbifolds possess a cyclic permutation symmetry.
Nevertheless, 
some of the vacua that may seem identical in
the low energy effective field theory limit
of the observable sector,
differ by other properties, such as the massive spectrum,  
superpotential couplings,
hidden sector matter states 
and are therefore distinct. 

\section{String Spectrum}\label{analysisspec}
The vector bosons from the untwisted sector generate the
\begin{equation*}
SU(5)\times U(1)\times{U(1)}^3\times SU(4) \times U(1) \times U(1) \times SO(6)
\end{equation*}

\noindent gauge symmetry.
Depending on the  choices of the GGSO projection coefficients,
extra space--time vector bosons may be obtained
from the following twelve sectors

\begin{equation}
\mathbf{G} =
\left\{ \begin{array}{ccccc}
\,\,\,\, z_1          ,&
\,\,\,\, z_2          ,&
\,\,\,\, z_1 + z_2    ,&
\,\,\,\, z_1 + 2\alpha,\\
\,\,\,\,\, \alpha       ,&
\,\,\,\,\, z_1 + \alpha ,&
\,\,\, z_2 + \alpha ,&
\,\, z_1 + z_2 + \alpha,\\
\,\, 3\alpha       ,&
\,\,\,\,\,\,\,\, z_1 + 3\alpha ,&
\,\,\,\,\,\, z_2 + 3\alpha ,&
\,\,\, z_1 + z_2 + 3\alpha
\end{array} \right\}. \label{ggsectors1}
\end{equation}
\\
The projections on the sectors
$3\alpha$, $z_1 + 3\alpha$, $z_2 + 3\alpha$, $z_1 + z_2 + 3\alpha$
can be inferred from the projections on the sectors
$\alpha$, $z_1 + \alpha$, $z_2 + \alpha$, $z_1 + z_2 + \alpha$
respectively. Therefore, we will not discuss them in detail. The
gauge bosons that are obtained from the sectors in (\ref{ggsectors1})
enhance the untwisted gauge symmetry.
We impose the restriction that the only gauge bosons that
remain in the spectrum are those that are obtained
from the untwisted sector.
The gauge groups in these models are therefore
\begin{eqnarray}
{\rm Observable} &: &~~~~SU(5)_{obs} \times U(1)_{5} 
                         \times U(1)_1 \times U(1)_2 \times U(1)_3 \nonumber\\
{\rm Hidden}     &: &~~~~SU(4)_{hid} \times U(1)_{4} \times U(1)_{hid} 
                                     \times SO(6)_{hid}             \nonumber
\end{eqnarray}
The NS sector matter spectrum 
is common in these models and consists of
three pairs of $\textbf 5$ and
$\overline{\textbf5}$ representations of the observable
$SU(5)\times U(1)_5$ gauge group
and twelve that are singlets under
the non--Abelian gauge symmetries.

\subsection{Observable Matter Spectrum}\label{analysis2}
The chiral matter spectrum arises from the twisted sectors.
The method of classification enables a straightforward
enumeration of all the twisted sectors that produce massless states
and the GGSO projection that operate on them. We provide below the
details of the method in the case of $\alpha_1$ in (\ref{alphas}).
The chiral spinorial representations of the observable $SU(5) \times U(1)_5$
arise from the sectors
\begin{eqnarray} \label{obspin}
B_{pqrs}^{(1)}&=& S + {b_1 + p e_3+ q e_4 + r e_5 + s e_6} \nonumber\\
&=&\{\psi^\mu,\chi^{12},(1-p)y^{3}\overline{y}^3,p\omega^{3}\overline{\omega}^3,
(1-q)y^{4}\overline{y}^4,q\omega^{4}\overline{\omega}^4, \nonumber\\
& & ~~~(1-r)y^{5}\overline{y}^5,r\omega^{5}\overline{\omega}^5,
(1-s)y^{6}\overline{y}^6,s\omega^{6}\overline{\omega}^6,
\overline{\eta}^1,\overline{\psi}^{1,...,5}\},
\\
B_{pqrs}^{(2)}&=& S + {b_2 + p e_1+ q e_2 + r e_5 + s e_6},
\label{twochiralspinorials}
\nonumber\\
B_{pqrs}^{(3)}&=& S + {b_3 + p e_1+ q e_2 + r e_3 + s e_4}, \nonumber
\end{eqnarray}
where $p,q,r,s=0,1$ and $b_3=b_1+b_2+2\alpha+z_1$.
These 48 sectors give rise to $\textbf{16}$ and $\overline{\textbf{16}}$ 
multiplets of $SO(10)$ decomposed under
$SU(5) \times U(1)$, which are given by
\begin{eqnarray}
\textbf{16} &= &\left(\overline{\textbf{5}},
-{\tfrac{{3}}{{2}}}\right) + 
\left(\textbf{10},+{\tfrac{{1}}{{2}}}\right) + 
\left(\textbf{1},+{\tfrac{{5}}{{2}}}\right),\nonumber\\
\overline{\textbf{16}} &= &\left(\textbf{5},
+{\tfrac{{3}}{{2}}}\right) + 
\left(\overline{\textbf{10}},-{\tfrac{\textbf{1}}{\textbf{2}}}\right)
+ \left(\textbf{1},-{\tfrac{{5}}{{2}}}\right).\nonumber
\end{eqnarray}
Additionally, vector--like representations of the observable
$SU(5) \times U(1)_5$ gauge group arise from the sectors
\begin{eqnarray}
B_{pqrs}^{(4)}&=& B_{pqrs}^{(1)} + z_1 + 2 \alpha
\nonumber\\
&=&\{\psi^\mu,\chi^{12},(1-p)y^{3}\overline{y}^3,p\omega^{3}\overline{\omega}^3,
(1-q)y^{4}\overline{y}^4,q\omega^{4}\overline{\omega}^4, \nonumber\\
& & ~~~~~~~~~(1-r)y^{5}\overline{y}^5,r\omega^{5}\overline{\omega}^5,
(1-s)y^{6}\overline{y}^6,s\omega^{6}\overline{\omega}^6,\overline{\eta}^{2,3} \},
\label{nonchiralvectorials}\\
B_{pqrs}^{(5,6)}&=& B_{pqrs}^{(2,3)} + z_1 + 2 \alpha. \nonumber
\end{eqnarray}

These sectors contain four periodic worldsheet right--moving complex 
fermions. The massless
states are obtained by acting on the vacuum with a Neveu--Schwarz
right--moving fermionic oscillator. Furthermore, if the oscillator is 
given by $\{\overline{\psi}^{1,\dots,5}\}$ or $\{\overline{\psi}^{*1,\dots,5}\}$ 
then some of the 48 twisted 
sectors can give rise to the vectorial \textbf{10} representation
of $SO(10)$ decomposed under $SU(5) \times U(1)$, which is given by
\begin{eqnarray}
\textbf{10} &= &\left(\overline{\textbf{5}},
+{{1}}\right) + 
\left(\textbf{5},-{{1}}\right). \nonumber
\end{eqnarray}
These states are identified with light Higgs 
representations that are used to break the Standard Model gauge 
symmetry to $SU(3)\times U(1)_{\rm e.m.}$. 
Additional states which are singlets under the observable $SU(5)\times U(1)_5$ 
might also arise from any of the 48 sectors in (\ref{nonchiralvectorials}), given by the following representations:

\begin{itemize}
\item $\{\overline\eta^{i}\}|R \rangle_{pqrs}^{(4,5,6)}$ or
$\{\overline\eta^{*i}\}|R\rangle_{pqrs}^{(4,5,6)}$, $i = 1,2,3$,
where $|R\rangle_{pqrs}^{(4,5,6)}$ is the degenerated Ramond vacuum of the
$B_{pqrs}^{(4,5,6)}$ sector.
These states transform as a vector--like representations under the $U(1)_i$'s.

\item $\{\overline\phi^{1,...,4}\}|R\rangle_{pqrs}^{(4,5,6)}$ or
$\{\overline\phi^{*1,...,4}\}|R\rangle_{pqrs}^{(4,5,6)}$.
These states transform as a vector--like representations 
of $SU(4) \times U(1)_4$.

\item $\{\overline\phi^{5}\}|R\rangle_{pqrs}^{(4,5,6)}$ or
$\{\overline\phi^{*5}\}|R\rangle_{pqrs}^{(4,5,6)}$.
These states transform as a vector--like representations under the $U(1)_5$'s.

\item $\{\overline\phi^{6,7,8}\}|R\rangle_{pqrs}^{(4,5,6)}$ or
$\{\overline\phi^{*6,7,8}\}|R\rangle_{pqrs}^{(4,5,6)}$.
These states transform as a vectorial representation of $SO(6)$.
\end{itemize}

\subsection{Hidden Matter Spectrum}\label{analysis3}
The sectors which produce states that transform under representations of
the hidden gauge group are singlets of the observable $SO(10)$ GUT gauge group.
These states are hidden 
matter states that are obtained 
in the string model, but are not
exotic with respect to the Standard Model gauge charges.
The 48 sectors in $B_{pqrs}^{1,2,3}+2\alpha$ produce states that transforms under the
$\left(\overline{\textbf{4}},+{{1}}\right)$, 
$\left(\textbf{4},-{{1}}\right)$, 
$\left(\textbf{6},{{0}}\right)$, 
$\left({\textbf{1}},+{{2}}\right)$ and 
$\left(\textbf{1},-{{2}}\right)$
representations of the $SU(4) \times U(1)$ hidden gauge group and are given by
\begin{eqnarray}\label{hidspin2}
B_{pqrs}^{(7)} &=& B_{pqrs}^{(1)} + 2\alpha
\nonumber\\
&=&\{\psi^\mu,\chi^{12},(1-p)y^{3}\overline{y}^3,p\omega^{3}\overline{\omega}^3,
(1-q)y^{4}\overline{y}^4,q\omega^{4}\overline{\omega}^4, \nonumber\\
& & ~~~(1-r)y^{5}\overline{y}^5,r\omega^{5}\overline{\omega}^5,
(1-s)y^{6}\overline{y}^6,s\omega^{6}\overline{\omega}^6,
\overline{\eta}^{2,3},\overline{\phi}^{1,...,4} \}, 
\\
B_{pqrs}^{(8, 9)} &=& B_{pqrs}^{(2, 3)} + 2\alpha. \nonumber
\end{eqnarray}
In addition we have the following 48 sectors
\begin{eqnarray}\label{hidspin3}
B_{pqrs}^{(10)} &=& B_{pqrs}^{(1)} + z_1 + z_2 + 2\alpha
\nonumber\\
&=&\{\psi^\mu,\chi^{12},(1-p)y^{3}\overline{y}^3,p\omega^{3}\overline{\omega}^3,
(1-q)y^{4}\overline{y}^4,q\omega^{4}\overline{\omega}^4, \nonumber\\
& & ~~~(1-r)y^{5}\overline{y}^5,r\omega^{5}\overline{\omega}^5,
(1-s)y^{6}\overline{y}^6,s\omega^{6}\overline{\omega}^6,
\overline{\eta}^{2,3},\overline{\phi}^{5,...,8} \},
 \\
B_{pqrs}^{(11, 12)} &=& B_{pqrs}^{(2, 3)} + z_1 + z_2 + 2\alpha, \nonumber
\end{eqnarray}
which produce states in the $\textbf{4}$ and $\overline{\textbf{4}}$ 
spinorial representations of the hidden $SO(6)$ gauge group. 

\subsection{Exotic Matter Spectrum}\label{analysis4}
In the string spectrum, additional sectors exist which produce 
fractionally charged states under the $SU(5) \times U(1)$ symmetry. These sectors arise when we have massless states, which are produced from a linear combination of basis vectors that include the vector $\alpha$, resulting in the breaking of $SO(10)$ symmetry. Moreover, these sectors produce states
that do not fall into representations of the underlying 
$SO(10)$ GUT symmetry. Specifically, they possess fractionally
charged assignments with respect to the $U(1)$ symmetry
in the decomposition $SO(10)\longrightarrow SU(5)\times U(1)$. 
Consequently, provided that the weak hypercharge has 
the canonical $SO(10)$ GUT embedding and the canonical
GUT prediction $\sin^2\theta_w=3/8$, these sectors produce
states that carry fractional electric charges. 
This is a generic feature of string compactifications
\cite{WW1985, Schellekens1990}, that may have interesting phenomenological
implications \cite{FC1992}, as electric charge 
conservation implies that the lightest of those exotic states is necessarily stable.
Many experimental searches for fractionally
charged matter have been conducted \cite{Halyo2000}.
However, no reported observation of any such particles has
ever been confirmed and there
are strong upper bounds on their abundance \cite{Halyo2000}.
This implies that such exotic states in string
models should be either confined into integrally
charged states \cite{RevampAEHN}, or be
sufficiently heavy and diluted in the cosmological
evolution of the universe \cite{FC1992}.
The first of these solutions is problematic, due to the effect of the
charged states on the renormalisation group running of the weak--hypercharge
and gauge coupling unification. The preferred solution is therefore for
the fractionally charged states to become sufficiently massive, {\it i.e.}
with a mass which is larger than the GUT scale. In this case the fractionally
charged states can be diluted by the inflationary evolution of the universe.
Due to their heavy mass they will not be reproduced during reheating and
the experimental constraints can be evaded.
Three generation Pati--Salam heterotic--string models in which the
fractionally charged states arise in the massive string
spectrum but not as massless states, that were constructed 
in \cite{ACFKR2011}, which are dubbed as the quasi--realistic 
exophobic Pati--Salam string models.
A particular question of interest in the current work
is the existence of quasi--realistic flipped $SU(5)$
heterotic--string models. Also, it should be noted that the 
sectors appearing in
(\ref{hidspin2}) and (\ref{hidspin3}) contain the 
combination $2\alpha$ and do not break the $SO(10)$ symmetry. Therefore,
these sectors do not produce exotic states under the $SU(5)\times U(1)$ gauge symmetry. 

In the free fermionic construction, we classify the sectors that produce exotic states according to
the product $\xi_R\cdot\xi_R=4$, 6, or 8. In the first case, 
massless states are obtained by acting on the vacuum with a 
Neveu--Schwarz fermion or with two oscillators with 1/4 frequencies.
In the second case, oscillators with 1/4 frequency are needed
to produce massless states, whereas in the third case no
oscillators are used to produce massless states. Furthermore, in the third case with no oscillators, we have the following 96 sectors
\begin{eqnarray}\label{exoticspin1}
B_{pqrs}^{(13)} &=& B_{pqrs}^{(1)} + z_2 + \alpha
\nonumber\\
&=&
\{\psi^\mu,\chi^{12},(1-p)y^{3}\overline{y}^3,p\omega^{3}\overline{\omega}^3,
(1-q)y^{4}\overline{y}^4,q\omega^{4}\overline{\omega}^4,
\nonumber\\
& & ~~~(1-r)y^{5}\overline{y}^5, r\omega^{5}\overline{\omega}^5,
(1-s)y^{6}\overline{y}^6,s\omega^{6}\overline{\omega}^6,\overline{\eta}^1 = 
- \textstyle\frac{1}{2},
\\
& & ~~~~~~~~~~~~~~~~~~
 \overline{\eta}^{2,3} = \textstyle\frac{1}{2}, \overline{\psi}^{1,...,5} = 
- \textstyle\frac{1}{2}, \overline{\phi}^{1,...,4} = 
\textstyle\frac{1}{2}, \overline{\phi}^{6,7,8} \}, \nonumber\\
B_{pqrs}^{(14 , 15)} &=& B_{pqrs}^{(2, 3)} + z_2  + \alpha, \nonumber
\end{eqnarray}

\begin{eqnarray}\label{exoticspin2}
B_{pqrs}^{(16)} &=& B_{pqrs}^{(1)} + z_1 + z_2  + \alpha \nonumber\\
&=&\{\psi^\mu,\chi^{12},(1-p)y^{3}\overline{y}^3,p\omega^{3}\overline{\omega}^3,
(1-q)y^{4}\overline{y}^4,q\omega^{4}\overline{\omega}^4,
\nonumber\\
& & ~~~r (1-r)y^{5}\overline{y}^5, \omega^{5}\overline{\omega}^5,
(1-s)y^{6}\overline{y}^6,s\omega^{6}\overline{\omega}^6,\overline{\eta}^1 = 
- \textstyle\frac{1}{2},
\\
& & ~~~~~~~~~~~~~~~~~~ 
\overline{\eta}^{2,3} = \textstyle\frac{1}{2},\overline{\psi}^{1,...,5} =  
- \textstyle\frac{1}{2}, \overline{\phi}^{1,...,4} = - \textstyle\frac{1}{2}, 
\overline{\phi}^{6,7,8} \},
\nonumber\\
B_{pqrs}^{(17 , 18)} &=& B_{pqrs}^{(2, 3)} + z_1 + z_2  + \alpha. \nonumber
\end{eqnarray}
These produce states that are singlets under the observable $SU(5)$ but
are charged under the $U(1)_5$ and are  given by
$\left({\textbf{1}},-\tfrac{{5}}{{4}}\right)$ 
and $\left(\textbf{1},+\tfrac{{5}}{{4}}\right)$. We now move on to the second case that consists of oscillators with one 1/4 frequency giving rise to 
additional massless
vector--like states given by the following 48 sectors
\begin{eqnarray}\label{vecto}
B_{pqrs}^{(19)} &=& B_{pqrs}^{(1)} + \alpha
\nonumber\\
&=&\{\psi^\mu,\chi^{12},(1-p)y^{3}\overline{y}^3,p\omega^{3}\overline{\omega}^3,
(1-q)y^{4}\overline{y}^4,q\omega^{4}\overline{\omega}^4,\nonumber\\
& & ~~~r (1-r)y^{5}\overline{y}^5, \omega^{5}\overline{\omega}^5,
(1-s)y^{6}\overline{y}^6,s\omega^{6}\overline{\omega}^6,\overline{\eta}^1 = 
- \textstyle\frac{1}{2},
\\
& & ~~~~~~~~~~~~~~~~~~~~~~~~
\overline{\eta}^{2,3} = \textstyle\frac{1}{2},\overline{\psi}^{1,...,5} =
- \textstyle\frac{1}{2}, \overline{\phi}^{1,...,4} = 
\textstyle\frac{1}{2}, \overline{\phi}^5 \},
\nonumber\\
B_{pqrs}^{(20, 21)} &=& B_{pqrs}^{(2, 3)} + \alpha. \nonumber
\end{eqnarray}
As an example, the sectors in $B^{(19)}_{pqrs}$ produce the following states:
\begin{itemize}
\item $\{\overline\eta^{1}\}|R\rangle_{pqrs}^{(19)}$,
where
$|R\rangle_{pqrs}^{(19)}$ is the degenerate 
Ramond vacuum of the 
$B_{pqrs}^{(19)}$ sector.
These states transform as vector--like
representations under the $U(1)_1$.

\item $\{\overline\eta^{*2}\}|R\rangle_{pqrs}^{(19)}$ and
$\{\overline\eta^{*3}\}|R\rangle_{pqrs}^{(19)}$.
These states transform as vector--like
representations under the $U(1)_{2/3}$.

\item $\{\overline{\psi}^{1,...,5}\}|R\rangle_{pqrs}^{(19)}$.
These states transform as
$\left(\overline{\textbf{5}},+\tfrac{{1}}{{4}}\right)$ 
and $\left(\textbf{5},-\tfrac{{1}}{{4}}\right)$
representations of $SU(5) \times U(1)$.

\item $\{\overline\phi^{*1,...,4}\}|R\rangle_{pqrs}^{(19)}$.
These states transform as vector--like 
representations of $SU(4) \times U(1)$.
\end{itemize}
Similarly the sectors in $B^{(20)}_{pqrs}$ and $B^{(21)}_{pqrs}$ produce the states above. What is more, similar states appear in the following 48 sectors
\begin{eqnarray}\label{bz1alpha}
B_{pqrs}^{(22)}&=& B_{pqrs}^{(1)} + z_1 + \alpha
\nonumber\\
&=&\{\psi^\mu,\chi^{12},(1-p)y^{3}\overline{y}^3,p\omega^{3}\overline{\omega}^3,
(1-q)y^{4}\overline{y}^4,q\omega^{4}\overline{\omega}^4,\nonumber\\
& & ~~~r (1-r)y^{5}\overline{y}^5, \omega^{5}\overline{\omega}^5,
(1-s)y^{6}\overline{y}^6,s\omega^{6}\overline{\omega}^6,\overline{\eta}^1 =
-\textstyle\frac{1}{2}, 
\\
& & ~~~~~~~~~~~~~~~~~~~~~
\overline{\eta}^{2,3} = \textstyle\frac{1}{2},\overline{\psi}^{1,...,5} = 
-\textstyle\frac{1}{2}, \overline{\phi}^{1,...,4} = 
-\textstyle\frac{1}{2}, \overline{\phi}^5 \},
\nonumber\\
B_{pqrs}^{(23,24)}&=& B_{pqrs}^{(2,3)} + z_1 + \alpha. \nonumber
\end{eqnarray}
The only difference between the sectors in (\ref{vecto})
and (\ref{bz1alpha}) is the sign of the $1/2$ boundary condition
of the worldsheet fermion ${\overline\phi}^{1,...,4}$. This changes
some of the $U(1)$ charges arising in (\ref{vecto}) compared 
to those arising in (\ref{bz1alpha}), but the structure and type 
of states are similar to those listed above. Finally, the 
first case of exotic states arise in the sectors
$\alpha$ and $z_1+\alpha$. These exotic states can be eliminated by 
the same conditions that eliminate the space--time 
vector bosons arising in these sectors which will be discussed in section \ref{gge}. 

\section{Twisted matter spectrum}\label{projectors}
The counting of spinorial and vector--like representations
in the given string vacua is realised by utilising the so
called projectors. Each sector $B^{i}_{pqrs}$, corresponds to a projector,
$P^{i}_{pqrs} = 0,1$, which is expressed in terms of GGSO coefficients
and determines whether a given sector survives the GGSO projections. 
It is noted with the basis vectors given in (\ref{basis}), 
each fixed point of the ${\mathbb{Z}}_2 \times {\mathbb{Z}}_2$ orbifold corresponds to 
a distinct sector $\xi$ in the additive group. In this method, the states
arising from each fixed point are, therefore, controlled individually. 
Furthermore, the computational analysis is facilitated
by rewriting the projectors in an analytic form. These are 
written as algebraic conditions, for the individual states
arising in the string spectrum, in terms of the GGSO phases 
of the basis vectors. The algebraic expressions are inserted 
into the computer code, which enables the scan of the large 
space of models spanned by the basis GGSO phases.

\subsection{Observable spinorial states}
In order to get the particle content for the representations for the sectors
in \eqref{obspin}, we used the following normalisations for the
hypercharge and the electromagnetic charge:
\begin{eqnarray}
Y &=& \frac{1}{3} (Q_1 + Q_2 + Q_3) + \frac{1}{2} (Q_4 + Q_5), \nonumber\\
Q_{em} &=& Y + \frac{1}{2} (Q_4 - Q_5). \nonumber
\end{eqnarray}
Where the $Q_{i}$ charges of a state, arise due to $\psi^{i}$ for $i =
1,...,5$. 
The following table summarises the charges of the colour $SU(3)$ and
electroweak $SU(2) \times U(1)$ Cartan generators, 
of the states which form the $SU(5) \times
U(1)$ matter representations

\begin{center}
\begin{tabular}{|c|c|c|c|c|}
\hline
 Representation & $\overline{\psi}^{1,2,3}$ &
$\overline{\psi}^{4,5}$ & $Y$ & $Q_{em}$ \\
\hline \hline
& $(+,+,+)$ & ($+,-$)& 1/2& 1,0\\
$\left( \, \textbf{5} \, , \, +\frac{{3}}{{2}} \, \right)$ & ($+,+,-$)& $(+,+)$ & 2/3& 2/3\\ 
\hline
& ($+,-,-$)& $(-,-)$ & -2/3& -2/3\\
$\left( \, \overline{\textbf{5}} \, , \, -\frac{{3}}{{2}} \, \right)$ & $(-,-,-)$ & ($+,-$)& -1/2& -1,0\\ 
\hline
& $(+,+,+)$ & $(-,-)$ & 0& 0\\
$\left(\textbf{10},+\frac{{1}}{{2}}\right)$& ($+,-,-$)& $(+,+)$ & 1/3& 1/3\\
& ($+,+,-$)& ($+,-$)& 1/6& -1/3,2/3\\ 
\hline
& ($+,+,-$)&
$(-,-)$ & -1/3& -1/3\\
$\left(\overline{\textbf{10}},-\frac{{1}}{{2}}\right)$ & ($+,-,-$)& ($+,-$)& -1/6& 1/3,-2/3\\
& $(-,-,-)$ & $(+,+)$ & 0& 0\\ 
\hline
$( \, \textbf{1} \, , \, +\frac{{5}}{{2}} \, )$&
$(+,+,+)$ & $(+,+)$ & 1& 1\\ 
\hline
$( \, \textbf{1} \, , \, -\frac{{5}}{{2}} \, )$
& $(-,-,-)$ & $(-,-)$ & -1& -1\\ 
\hline
\end{tabular}
\end{center}
Here $``+"$, and $``-"$,  label the contribution of an
oscillator with fermion number $F = 0$, or $F = -1$,  to the degenerate vacuum.
For example $(+,+,-)$ under $\overline{\psi}^{1,2,3}$ corresponds
to a part of the Ramond vacuum formed by two oscillators with fermion number
$F = 0$ and one oscillator with fermion numbers $F = -1$.
These states correspond to particles of the Standard Model.
More precisely we can decompose these representations under
$SU(3) \times SU(2) \times U(1)$:
\begin{align} \label{16decomposition}
\left( \,  \,  \overline{\textbf{5}} \,  ,-\frac{3}{2}\right)&
=\left(\overline{\textbf{3}},\textbf{1},-\frac{2}{3}\right)_{u^c}+\left(\textbf{1},\textbf{2},-\frac{1}{2}\right)_{L}, \nonumber \\
\left(\textbf{10},+\frac{1}{2}\right)&=\left(\textbf{3},\textbf{2},+\frac{1}{6}\right)_{Q} \, 
+\left(\overline{\textbf{3}},\textbf{1},+\frac{1}{3}\right)_{d^c}+\left(\textbf{1},\textbf{1},0\right)_{\nu^c},\nonumber\\
\left( \,  \,  \textbf{1} \, ,+\frac{5}{2}\right)&=\left(\textbf{1},\textbf{1},+1 \, \right)_{e^c}, \nonumber
\end{align}
where $L$ is the lepton--doublet; $Q$ is the quark--doublet; 
$d^c,~u^c,~e^c$ and $\nu^c$ are the quark and lepton singlets.
Because of the $\alpha$--projection, which projects on incomplete
$\textbf{16}$ and $\overline{\textbf{16}}$ representations, complete families and 
anti--families are formed by combining states from different sectors. 

\subsection{Chirality Operators}\label{chiralityoperators}
A phenomenologically viable model consists of 3 families of 
chiral $\textbf{16}$ representations of $SO(10)$ decomposed
under $SU(5)\times U(1)$. 
Therefore, we have to count the number of $\textbf{16}$s and
$\overline{\textbf{16}}$s. The choice of GGSO coefficients
determine the model we consider and therefore the number of families.
In order to be able to distinguish between $\textbf{16}$ and
$\overline{\textbf{16}}$, one has to define operators that
determine the representations in which the states of each observable
sector fall into.
The operators $X_{pqrs}^{{(1,2,3)}_{SO(10)}} = \pm 1$, defines the $SO(10)$ chirality $(\mathbf{16}$ or $\mathbf{\overline{16}})$
for $B^{1}_{pqrs}$, $B^{2}_{pqrs}$ and $B^{3}_{pqrs}$, which are given by
\begin{eqnarray}\label{so10operators}
X_{pqrs}^{(1)_{SO(10)}} & = &
C\binom{B^{(1)}_{pqrs}}{b_{2} + (1-r)e_{5} + (1-s)e_{6}},\nonumber\\
X_{pqrs}^{(2)_{SO(10)}} & = &
C\binom{B^{(2)}_{pqrs}}{b_{1} + (1-r)e_{5} + (1-s)e_{6}},\\
X_{pqrs}^{(3)_{SO(10)}} & = &
C\binom{B^{(3)}_{pqrs}}{b_{1} + (1-r)e_{3} + (1-s)e_{4}}.\nonumber
\end{eqnarray}
This is in contrast to
the case in the Pati--Salam heterotic--string models, where
one needs to determine the chirality of the $SO(6)$ and $SO(4)$
representations separately \cite{ACFKR2011}. 
Additionally, we determine which components in the $\textbf{16}$ and 
$\overline{\textbf{16}}$ survive the $\alpha$ projection, which breaks $SO(10)$ to $SU(5)\times U(1)$.
In this respect, we note that the $\alpha$ projection operates
identically on the ${\textbf{1}}\equiv({\bf1},+5/2)$ and 
${\overline{\textbf5}}\equiv({\overline{\textbf5}},-3/2)$
states and similarly on the conjugate representations
${\overline{\textbf{1}}}\equiv({\bf1},-5/2)$ 
and ${\textbf{5}}\equiv({\textbf5},+3/2)$. 
The surviving components are determined
by defining the operators 
$X_{pqrs}^{{(1,2,3)}_{SU(5)}} = \pm 1$,
where $X_{pqrs}^{{(i)}_{SU(5)}} = 1$ indicates survival of the 
$({\bf 1},+5/2)$ and $(\overline{\textbf5}, -3/2)$ pair 
and $X_{pqrs}^{{(1,2,3)}_{SU(5)}} = - 1$ indicates survival of 
the $(\textbf{10}, +1/2)$ states. The operator $X_{pqrs}^{{(i)}_{SU(5)}}$ acts similarly on the $\overline{\textbf{16}}$ of $SO(10)$.
These conditions are expressed as 
\begin{eqnarray}\label{su5operators}
X_{pqrs}^{(1)_{SU(5)}} & = &
C\binom{B^{(1)}_{pqrs}}{\alpha},\nonumber\\
X_{pqrs}^{(2)_{SU(5)}} & = &
C\binom{B^{(2)}_{pqrs}}{\alpha},\\
X_{pqrs}^{(3)_{SU(5)}} & = &
C\binom{B^{(3)}_{pqrs}}{\alpha}.\nonumber
\end{eqnarray} 

\subsection{Projectors}
The states in the sectors in $B_{pqrs}^{(1)}$, as given in (\ref{obspin}), can be projected in or out of the
string spectrum depending on the GGSO projections of the vectors $e_1$, $e_2$, $z_1$ and
$z_2$. Likewise for $B_{pqrs}^{(2)}$ and $B_{pqrs}^{(3)}$, we define a projector $P$ such that
the states survive when $P=1$ and are projected out when $P=0$, which are given as:

\footnotesize
\begin{align}
P_{pqrs}^{(1)} &= \frac{1}{16} 
\left( 1-C \binom {e_1} {B_{pqrs}^{(1)}}\right) . 
\left( 1-C \binom {e_2} {B_{pqrs}^{(1)}}\right) . 
\left( 1-C \binom {z_1} {B_{pqrs}^{(1)}}\right) . 
\left( 1-C \binom {z_2} {B_{pqrs}^{(1)}}\right),\nonumber\\
P_{pqrs}^{(2)} &= \frac{1}{16} 
\left( 1-C \binom {e_3} {B_{pqrs}^{(2)}}\right) . 
\left( 1-C \binom {e_4} {B_{pqrs}^{(2)}}\right) . 
\left( 1-C \binom {z_1} {B_{pqrs}^{(2)}}\right) . 
\left( 1-C \binom {z_2} {B_{pqrs}^{(2)}}\right),\\
P_{pqrs}^{(3)} &= \frac{1}{16} 
\left( 1-C \binom {e_5} {B_{pqrs}^{(3)}}\right) . 
\left( 1-C \binom {e_6} {B_{pqrs}^{(3)}}\right) . 
\left( 1-C \binom {z_1} {B_{pqrs}^{(3)}}\right) . 
\left( 1-C \binom {z_2} {B_{pqrs}^{(3)}}\right).\nonumber
\end{align}
\normalsize

These projectors can be expressed as a system of linear equations with $p$,
$q$, $r$ and $s$ as unknowns. The solutions of such a system of equations
yield the different combinations of $p$, $q$, $r$ and $s$ for which sectors
survive the GGSO projections. The analytic expressions for each of the different
projectors $P^{1,2,3}_{pqrs}$ are given in a matrix form
$\Delta^{i}W^{i} = Y^{i}$, where

\begin{align}
\begin{pmatrix} (e_1|e_3)&(e_1|e_4)&(e_1|e_5)&(e_1|e_6)\\
(e_2|e_3)&(e_2|e_4)&(e_2|e_5)&(e_2|e_6)\\
(z_1|e_3)&(z_1|e_4)&(z_1|e_5)&(z_1|e_6)\\
(z_2|e_3)&(z_2|e_4)&(z_2|e_5)&(z_2|e_6) \end{pmatrix}
\begin{pmatrix} p\\q\\r\\s\end{pmatrix} &=
\begin{pmatrix} (e_1|b_1)\\
(e_2|b_1)\\
(z_1|b_1)\\
(z_2|b_1)
\end{pmatrix}=Y^1,\nonumber
\\[0.3cm]
\begin{pmatrix} (e_3|e_1)&(e_3|e_2)&(e_3|e_5)&(e_3|e_6)\\
(e_4|e_1)&(e_4|e_2)&(e_4|e_5)&(e_4|e_6)\\
(z_1|e_1)&(z_1|e_2)&(z_1|e_5)&(z_1|e_6)\\
(z_2|e_1)&(z_2|e_2)&(z_2|e_5)&(z_2|e_6) \end{pmatrix}
\begin{pmatrix} p\\q\\r\\s\end{pmatrix} &=
\begin{pmatrix} (e_3|b_2)\\
(e_4|b_2)\\
(z_1|b_2)\\
(z_2|b_2)
\end{pmatrix}=Y^2,
\\[0.3cm]
\begin{pmatrix} (e_5|e_1)&(e_5|e_2)&(e_5|e_3)&(e_5|e_4)\\
(e_6|e_1)&(e_6|e_2)&(e_6|e_3)&(e_6|e_4)\\
(z_1|e_1)&(z_1|e_2)&(z_1|e_3)&(z_1|e_4)\\
(z_2|e_1)&(z_2|e_2)&(z_2|e_3)&(z_2|e_4) \end{pmatrix}
\begin{pmatrix} p\\q\\r\\s\end{pmatrix} &=
\begin{pmatrix} (e_5|b_3)\\
(e_6|b_3)\\
(z_1|b_3)\\
(z_2|b_3)
\end{pmatrix}=Y^3.\nonumber
\end{align}
Here the GGSO phases are defined as
$$C \binom{v_i}{v_j} = e^{i \pi (v_i|v_j)}~,$$
where $v_i$ and $v_j$ refer to the basis vectors and the GGSO
projections are defined as in (\ref{gso}).  
The corresponding algebraic expressions for the states from the 
remaining sectors
above are enumerated in the appendix, as well as the states in the hidden sector 
which are not classified here, may play a substantial role in the string phenomenology, such as in the case of the SUSY breaking. Furthermore, the projectors presented in the appendix
determine the number of surviving observable, hidden and exotic
states in each model.

\section{Gauge Group Enhancements}\label{gge}
The $SU(5)\times U(1)$ gauge symmetry generated by the
untwisted space--time vector bosons, may be enhanced by the 
vector bosons that arise from the sectors listed
in (\ref{ggsectors1}).
We impose that all the additional space--time vector bosons are 
projected out. The gauge symmetry is therefore identical in all the 
models that we scan, though the occurrence
of models with enhancements is approximately about 23.8\% of the total models.  
The string models in our classification differ by the 
string spectrum that arises from the twisted sectors.   
In our classification method, we encode the GGSO projections coefficients in terms of algebraic equations, which are applied to all the sectors listed in section
\ref{analysisspec}.

The gauge bosons of any given sector in (\ref{ggsectors1})
transform under a subgroup of the Neveu--Schwarz gauge group.
If they survive the GGSO projections,
then the NS gauge group is enhanced.
We restrict our classification here
to the cases without enhancement, by identifying when the gauge bosons survive the GGSO projections and generalize the formulae to eliminate them.
We remark that models with enhanced gauge symmetry 
in the observable or hidden sectors may be of interest
for various phenomenological reasons, as, for example, 
the $SU(6)\times SU(2)$ string models presented 
in \cite{SU6SU2}. 
Below, we present the different types of
enhancements that can occur within the string spectrum from the sectors given in (\ref{ggsectors1}). In addition, we assume that only one set of conditions is satisfied from any one given sector in (\ref{ggsectors1}).

\subsection{Observable gauge group enhancement}
There is one sector contributing only to the enhancement of the
observable gauge group i.e.
$SU(5)_{obs}\times U(1)_5 \times U(1)_1 \times U(1)_2 \times U(1)_3 $.
This is the sector $z_1 + 2 \alpha$, given by the conditions:
\begin{itemize}
\item
$z_1 + 2 \alpha = \{ \overline{\psi}^{1,...,5}, \overline{\eta}^{1,2,3} \}$
\begin{center}
   \begin{tabularx}{\textwidth}{|X|}
 
     \hline
     \textbf{Sector Condition}  \\ \hline
     $(z_1 + 2 \alpha|e_i) = (z_1 + 2 \alpha|z_k) = 0$
     \\ \hline
   \end{tabularx}
   
   \begin{tabularx}{\textwidth}{|X|X|}
 
     \hline
     \textbf{Enhancement Condition} &  \textbf{Resulting Enhancement} \\ \hline
$(z_1 + 2 \alpha|\alpha) \, = (z_1 + 2 \alpha|b_2)$ & 
$SU(5)_{obs}\times U(1)_5 \times U(1)_1 \times U(1)_2 \times U(1)_3 $
\newline
$ \xrightarrow{\hspace*{1cm}} SU(6) \times SU(2) \times U(1)^2$ \\ \hline
     $(z_1 + 2 \alpha|\alpha) \, \neq (z_1 + 2 \alpha|b_2)$ & 
$SU(5)_{obs}\times U(1)_5 \times U(1)_1 \times U(1)_2 \times U(1)_3 $
\newline$ \xrightarrow{\hspace*{1cm}} SO(10) \times U(1)^3$ \\ \hline
   \end{tabularx}
\end{center}
where $i= 1,\dots,6$ and $k=1,2$.

\end{itemize}
\newpage
\subsection{Hidden gauge group enhancement}
The vector bosons arising from the untwisted sector produce
the hidden gauge symmetry, which is given as
$SU(4)_{hid} \times U(1)_4 \times SO(6)_{hid} \times U(1)_{hid}$. 
Similar to the observable sector, there is one sector that enhances
only the untwisted hidden sector gauge symmetry and 
is given by the sector $z_1 + z_2$, where the conditions are given by:

\begin{itemize}

\item $z_1 + z_2 = \{ \overline{\phi}^{1,...,8}\}$ 
\begin{center}
    \begin{tabularx}{\textwidth}{|X|}
      \hline
      \textbf{Sector Condition}  \\ \hline
      $(z_1+z_2|e_i)= (z_1+z_2|b_k) =0$ 
      \\ \hline
    \end{tabularx}
    
   \begin{tabularx}{\textwidth}{|X|X|}
     \hline
     \textbf{Enhancement Condition} &  \textbf{Resulting Enhancement} \\ \hline
     $(z_1+z_2|z_1)= 1$ & $SU(4)_{hid} \times U(1)_4 \times SO(6)_{hid} \times U(1)_{hid}  $\newline$ \xrightarrow{\hspace*{1cm}} SU(8) \times U(1)$ \\ \hline
   \end{tabularx}
\end{center}
where $i= 1,\dots,6$ and $k=1,2$.

\end{itemize}

\subsection{Mixed gauge group enhancements}
The additional sectors in (\ref{ggsectors1}), produce
vector bosons coming from the mixture of the observable and hidden sector gauge groups. 
The mixed gauge group enhancements are formed from the untwisted
symmetries of the observable and hidden gauge group.
These are given from the sectors
$z_1$, $z_2$, $\alpha$, $z_1 + \alpha$, $z_2 + \alpha$ and
$z_1 + z_2 + \alpha$. The conditions are as follows:

\begin{itemize}

\item
$z_2 = \{ \overline{\phi}^{5,...,8}\}$
\begin{center}
   \begin{tabularx}{\textwidth}{|X|}
 
     \hline
     \textbf{Sector Condition}  \\ \hline
     $(z_2|e_i)= 0$
     \\ \hline
   \end{tabularx}
   
   \begin{tabularx}{\textwidth}{|X|X|}
 
     \hline
     \textbf{Enhancement Condition} &  \textbf{Resulting Enhancement} \\ 
\hline
     $(z_2|z_1)=1$ \newline $(z_2|b_k) =0$ & $SU(4)_{hid} \times U(1)_4 
                           \times SO(6)_{hid} \times U(1)_{hid} $\newline$ 
                          \xrightarrow{\hspace*{1cm}} SU(8) \times U(1)$ \\ 
\hline
     $(z_2|z_1) =0$ \newline $(z_2|b_k) \neq 1$ & $U(1)_{1/2/3} 
                           \times SO(6)_{hid} \times U(1)_{hid} $\newline$ 
                          \xrightarrow{\hspace*{1cm}} SU(5) \times U(1)$ \\    
\hline
     $(z_2|z_1) =0$ \newline $(z_2|b_k)=1$ & $SU(5)_{obs} \times U(1)_5 
                               \times SO(6)_{hid} \times U(1)_5$\newline$ 
                          \xrightarrow{\hspace*{1cm}} SU(9) \times U(1)$ \\     
\hline
   \end{tabularx}
 \end{center}
where $i= 1,\dots,6$ and $k=1,2$.

\newpage

\item $z_1 = \{ \overline{\phi}^{1,...,4} \}$
\begin{center}
   \begin{tabularx}{\textwidth}{|X|X|}
     \hline
     \textbf{Enhancement Condition} &  \textbf{Resulting Enhancement} \\ \hline
     $(z_1|e_i) = (z_1|b_k) = (z_1|z_1) = (z_1|\alpha) = 0$ \newline $(z_1|z_2) = 1$ & $SU(4)_{hid} \times U(1)_4 \times U(1)_{hid} \times SO(6)_{hid}$\newline$ \xrightarrow{\hspace*{1cm}} SO(10) \times SU(3) \times SU(2)$ \\ \hline
     $(z_1|e_i) = (z_1|b_k) = (z_1|z_1) = 0$ \newline $(z_1|z_2) = (z_1|\alpha) = 1$ & $SU(4)_{hid} \times U(1)_4 \times U(1)_{hid} \times SO(6)_{hid} $\newline$ \xrightarrow{\hspace*{1cm}} SO(12) \times SO(4)$ \\     \hline
     $(z_1|e_i)= (z_1|z_2) =0$ \newline $(z_1|b_k) \neq 1$ \newline $(z_1|z_1) = 1$ & $U(1)_{1/2/3} \times SU(4)_{hid} \times U(1)_4$\newline$ \xrightarrow{\hspace*{1cm}} SU(5) \times U(1)$ \\     \hline
     $(z_1|e_i) = (z_1|z_2) = 0$ \newline $(z_1|b_k) = (z_1|z_1) = 1$ & $SU(5)_{obs} \times U(1)_5 \times SU(4)_{hid} \times U(1)_4$\newline$ \xrightarrow{\hspace*{1cm}} SU(9) \times U(1)$ \\     \hline
     $(z_1|e_j) = (z_1|z_1) = (z_1|z_2) = (z_1|\alpha) = 0$ \newline $(z_1|e_i)=1$ \newline AND \newline $(z_1|b_1)=0$, $i=1,2$ \newline or \newline $(z_1|b_2)=0$, $i=3,4$ \newline or \newline $(z_1|b_1)=(z_1|b_2)$, $i=5,6$ & $SU(4)_{hid} \times U(1)_4 \xrightarrow{\hspace*{1cm}} SO(5) \times SO(5)$ \\     \hline
     $(z_1|e_j) = (z_1|z_1) = (z_1|z_2) = 0$ \newline $(z_1|e_i)=(z_1|\alpha)=1$ \newline AND \newline $(z_1|b_1)=0$, $i=1,2$ \newline or \newline $(z_1|b_2)=0$, $i=3,4$ \newline or \newline $(z_1|b_1)=(z_1|b_2)$, $i=5,6$ & $SU(4)_{hid} \times U(1)_4 \xrightarrow{\hspace*{1cm}} SO(8)$ \\     \hline
   \end{tabularx}
\end{center}
where $i,j= 1,\dots,6$, $i \neq j$ and $k=1,2$.

\newpage

\item $\alpha = \{ \overline{\psi}^{1,...,5}= \textstyle\frac{1}{2},
                   \overline{\eta}^{1,2,3}= \textstyle\frac{1}{2},
                   \overline{\phi}^{1,...,4} = \textstyle\frac{1}{2},
                   \overline{\phi}^{5} \}$
                   
The states that give two $\frac{1}{4}$ oscillators are produced from the following conditions:
\begin{center}
   \begin{tabularx}{\textwidth}{|X|}
     \hline
     \textbf{Sector Condition}  \\ \hline
     $(\alpha|e_i)= 0$ \newline $(\alpha|z_2) \neq (\alpha|\alpha) $
     \\ \hline
   \end{tabularx}
   
   \begin{tabularx}{\textwidth}{|X|X|}
 
     \hline
     \textbf{Enhancement Condition} &  \textbf{Resulting Enhancement} \\ \hline
     $(\alpha|z_1)= (\alpha|b_k)=0$ & $SU(5)_{obs} \times U(1)_5 \times U(1)_1 
       \times U(1)_2 \times U(1)_3 \times SU(4)_{hid} \times U(1)_4 
       \times U(1)_{hid}$\newline$ 
       \xrightarrow{\hspace*{1cm}} SO(10) \times SO(4) \times SO(4) 
       \times U(1)^4$ \\ \hline
     $(\alpha|z_1)=0$ \newline $(\alpha|b_k) \neq 0$ & $SU(5)_{obs} 
     \times U(1)_5 \times U(1)_1 \times U(1)_2 \times U(1)_3 
     \times U(1)_{hid} $\newline$ \xrightarrow{\hspace*{1cm}} SU(5) 
     \times SU(3) \times U(1)^3$ \\ \hline
     $(\alpha|z_1)=1$ \newline $(\alpha|b_k) \neq 1$ & $U(1)_{1/2/3} 
    \times SU(4)_{hid} \times U(1)_4 \times U(1)_{hid}$\newline$ 
    \xrightarrow{\hspace*{1cm}} SO(6) \times SO(4) \times U(1)$ \\ \hline
     $(\alpha|z_1)=(\alpha|b_k)=1$ & $SU(5)_{obs} \times U(1)_5 
     \times SU(4)_{hid} \times U(1)_4 \times U(1)_{hid} $\newline$ 
     \xrightarrow{\hspace*{1cm}} SU(7) \times SO(4) \times SU(3)$ \\ \hline
   \end{tabularx}
 \end{center}
where $i= 1,\dots, 6$ and $k=1,2$. Additionally, the states that give one $\frac{1}{2}$ oscillators are produced from the following conditions:
\begin{center}
   \begin{tabularx}{\textwidth}{|X|}
 
     \hline
     \textbf{Sector Condition}  \\ \hline
     $(\alpha|z_1)= 0$
     \\ \hline
   \end{tabularx}
   
   \begin{tabularx}{\textwidth}{|X|X|}
 
     \hline
     \textbf{Enhancement Condition} &  \textbf{Resulting Enhancement} \\ \hline
     $(\alpha|e_i)= (\alpha|b_k)=0$ \newline $(\alpha|z_2) \neq (\alpha|\alpha)$ & $SO(6)_{hid} \times U(1)_{hid} \xrightarrow{\hspace*{1cm}} SO(7) \times U(1)$ \\ \hline
          $(\alpha|z_2) = (\alpha|\alpha)$ \newline $(\alpha|e_j)=0$ \newline $(\alpha|e_i)=1$ \newline AND \newline $(\alpha|b_1)=0$, $i=1,2$ \newline or \newline $(\alpha|b_2)=0$, $i=3,4$ \newline or \newline $(\alpha|b_1)=(\alpha|b_2)$, $i=5,6$
          & $U(1)_{hid} \xrightarrow{\hspace*{1cm}} SU(2)$ \\ \hline

   \end{tabularx}
 \end{center}
where $i,j= 1,\dots,6$, $i \neq j$ and $k=1,2$.

\item $z_1 + \alpha = \{ \overline{\psi}^{1,...,5}= \textstyle\frac{1}{2}, 
                         \overline{\eta}^{1,2,3}= \textstyle\frac{1}{2},
                         \overline{\phi}^{1,...,4} = -\textstyle\frac{1}{2},
                         \overline{\phi}^{5} \}$

The states that give two $\frac{1}{4}$ oscillators are produced from the following conditions:
\begin{center}
   \begin{tabularx}{\textwidth}{|X|}
 
     \hline
     \textbf{Sector Condition}  \\ \hline
     $(z_1 + \alpha|e_i)= 0$
     \\ \hline
   \end{tabularx}
   
   \begin{tabularx}{\textwidth}{|X|X|}
 
     \hline
     \textbf{Enhancement Condition} &  \textbf{Resulting Enhancement} \\ \hline
     $(z_1 + \alpha|z_1)=(z_1 + \alpha|b_k)=0$ \newline $(z_1 + \alpha|z_2) \neq (z_1 + \alpha|\alpha)$ & $SU(5)_{obs} \times U(1)_5 \times U(1)_1 \times U(1)_2 \times U(1)_3 \times SU(4)_{hid} \times U(1)_4 \times U(1)_{hid}$\newline$ \xrightarrow{\hspace*{1cm}} SO(10) \times SO(4) \times SO(4) \times U(1)^4$ \\ \hline
     $(z_1 + \alpha|z_1)=0$ \newline $(z_1 + \alpha|b_k) \neq 0$ \newline $(z_1 + \alpha|z_2) \neq (z_1 + \alpha|\alpha)$ & $SU(5)_{obs} \times U(1)_5 \times U(1)_1 \times U(1)_2 \times U(1)_3 \times U(1)_{hid} $\newline$ \xrightarrow{\hspace*{1cm}} SU(5) \times SU(3) \times U(1)^3$ \\ \hline
     $(z_1 + \alpha|z_1)=1$ \newline $(z_1 + \alpha|b_k) \neq 1$  \newline $(z_1 + \alpha|z_2)=(z_1 + \alpha|\alpha)$ & $U(1)_{1/2/3} \times SU(4)_{hid} \times U(1)_4 \times U(1)_{hid}$\newline$ \xrightarrow{\hspace*{1cm}} SO(6) \times SO(4) \times U(1)$ \\ \hline
     $(z_1 + \alpha|z_1)=(z_1 + \alpha|b_k)=1$ \newline $(z_1 + \alpha|z_2)=(z_1 + \alpha|\alpha)$ & $SU(5)_{obs} \times U(1)_5 \times SU(4)_{hid} \times U(1)_4 \times U(1)_{hid} $\newline$ \xrightarrow{\hspace*{1cm}} SU(7) \times SO(4) \times SU(3)$ \\ \hline
   \end{tabularx}
 \end{center}
where $i= 1,\dots,6$ and $k=1,2$. Additionally, the states that give one $\frac{1}{2}$ oscillators are produced from the following conditions:
\begin{center}
   \begin{tabularx}{\textwidth}{|X|}
 
     \hline
     \textbf{Sector Condition}  \\ \hline
     $(z_1 + \alpha|z_1)= 0$      \\ \hline
   \end{tabularx}
   
   \begin{tabularx}{\textwidth}{|X|X|}
 
     \hline
     \textbf{Enhancement Condition} &  \textbf{Resulting Enhancement} \\ \hline
     $(z_1 + \alpha|e_i)= (z_1 + \alpha|b_k)=0$ \newline $(z_1 + \alpha|z_2) \neq (z_1 + \alpha|\alpha)$ & $SO(6)_{hid} \times U(1)_{hid} \xrightarrow{\hspace*{1cm}} SO(7) \times U(1)$ \\ \hline
     $(z_1 + \alpha|z_2)=(z_1 + \alpha|\alpha)$ \newline 
     $(z_1 + \alpha|e_j)=0$ \newline $(z_1 + \alpha|e_i)=1$ \newline 
     AND \newline $(z_1 + \alpha|b_1)=0$, $i=1,2$ \newline or \newline $(z_1 + \alpha|b_2)=0$, $i=3,4$ \newline or \newline $(z_1 + \alpha|b_1)=(z_1 + \alpha|b_2)$, $i=5,6$
     & $U(1)_{hid} \xrightarrow{\hspace*{1cm}} SU(2)$ \\ \hline

   \end{tabularx}
 \end{center}
where $i,j= 1,\dots,6$, $i \neq j$ and $k=1,2$.

\item $z_2 + \alpha = \{ \overline{\psi}^{1,..,5}= \textstyle\frac{1}{2},
                        \overline{\eta}^{1,2,3}= \textstyle\frac{1}{2},
                        \overline{\phi}^{1,..,4} = \textstyle\frac{1}{2},
                        \overline{\phi}^{6,7,8} \}$ \begin{center}
   \begin{tabularx}{\textwidth}{|X|}
    
        \hline
        \textbf{Sector Condition}  \\ \hline
        $(z_2 + \alpha|e_i)= 0$
        \newline $(z_2 + \alpha|\alpha)=\frac{1}{2}$
        \\ \hline
   \end{tabularx}
      
   \begin{tabularx}{\textwidth}{|X|X|}
 
     \hline
     \textbf{Enhancement Condition} &  \textbf{Resulting Enhancement} \\ \hline
     $(z_2 + \alpha|z_1)=0$  \newline $(z_2 + \alpha|b_k)=1$ & $SU(5)_{obs} \times U(1)_5 \times SO(6)_{hid} $\newline$ \xrightarrow{\hspace*{1cm}} SO(10) \times SO(6)$ \\ \hline
     $(z_2 + \alpha|z_1)=0$ \newline $(z_2 + \alpha|b_k) \neq 1$ & $U(1)_{1/2/3} \times SO(6)_{hid}$ \newline $ \xrightarrow{\hspace*{1cm}} SO(5) \times SO(5)$ \\ \hline
     $(z_2 + \alpha|b_k)=0$ \newline $(z_2 + \alpha|z_1)=1$ & $SU(4)_{hid} \times U(1)_4 \times SO(6)_{hid}$ \newline $ \xrightarrow{\hspace*{1cm}} SO(10) \times U(1)^2$ \\ \hline
   \end{tabularx}
 \end{center}
where $i= 1,\dots,6$ and $k=1,2$.

\item $z_1 + z_2 + \alpha = \{ \overline{\psi}^{1,..,5}= \textstyle\frac{1}{2}, 
                               \overline{\eta}^{1,2,3}= \textstyle\frac{1}{2}, 
                              \overline{\phi}^{1,..,4} = -\textstyle\frac{1}{2},
                              \overline{\phi}^{6,7,8} \}$
\begin{center}
   \begin{tabularx}{\textwidth}{|X|}
          
              \hline
              \textbf{Sector Condition}  \\ \hline
              $(z_1 + z_2 + \alpha|e_i)= 0$
              \\ \hline
   \end{tabularx}
      
   \begin{tabularx}{\textwidth}{|X|X|}
 
     \hline
     \textbf{Enhancement Condition} &  \textbf{Resulting Enhancement} \\ \hline
     $(z_1 + z_2 + \alpha|z_1)=0$  \newline $(z_1 + z_2 + \alpha|\alpha)=\frac{1}{2}$ \newline $(z_1 + z_2 + \alpha|b_k)=1$ & $SU(5)_{obs} \times U(1)_5 \times SO(6)_{hid}$  \newline $ \xrightarrow{\hspace*{1cm}} SO(10) \times SO(6)$ \\ \hline
     $(z_1 + z_2 + \alpha|z_1)=0$  \newline $(z_1 + z_2 + \alpha|\alpha)=\frac{1}{2}$ \newline $(z_1 + z_2 + \alpha|b_k) \neq 1$ & $U(1)_{1/2/3} \times SO(6)_{hid}$  \newline $ \xrightarrow{\hspace*{1cm}} SO(5) \times SO(5)$ \\ \hline
          $(z_1 + z_2 + \alpha|b_k)=0$  \newline $(z_1 + z_2 + \alpha|\alpha)=-\frac{1}{2}$ \newline $(z_1 + z_2 + \alpha|z_1)=1$ & $SU(4)_{hid} \times U(1)_4 \times SO(6)_{hid}$  \newline $ \xrightarrow{\hspace*{1cm}} SO(10) \times U(1)^2$ \\ \hline
   \end{tabularx}
 \end{center}
where $i= 1,\dots, 6$ and $k=1,2$.
 
\end{itemize}
Finally, we remark that as 
noted in section \ref{analysis4}, the sectors $\alpha$,
$z_1+\alpha$, $z_2+\alpha$ and $z_1+z_2+\alpha$ may also give rise to
exotic states, when the left--moving $\psi^\mu$ oscillator
is replaced by a left--moving $\chi^i$ oscillator. Moreover, we note that
the GGSO projections of the basis vectors $e_{1,...,6}$, $z_{1,2}$
and $\alpha$ do not 
distinguish between $\psi^\mu$ and $\chi^i$, which can therefore 
be used to project both the enhancements, as well as the exotic
states arising from the sectors $\alpha$, $z_1+\alpha$, $z_2+\alpha$
and $z_1+z_2+\alpha$. 

\section{Results}
By use of the algebraic expressions given in the sections previously, as well as in the appendix, we analyse
the entire massless spectrum for a given choice of configuration of
GGSO projection coefficients.
These expressions can be seen as matrix equations which we programmed into a JAVA and FORTRAN95 code independently, that are used 
to scan the space of the string vacua. 
The number of possible configurations is $2^{44}\approx 10^{13}$,
which is very large, for a
classification of the entire string vacua. For this purpose, a random generation
algorithm is
used\footnote{We, herein acknowledge the analysis of large sets
of string vacua which has been performed by other groups \cite{StatStudy1989}.}
and the characteristics of the models for each set of
random GGSO projection
coefficients are extracted. From the generated sample, a model
with the desired phenomenological
criteria can be fished out. This procedure was followed in \cite{ACFKR2011, CFR2011},
which produced a three generation Pati--Salam
heterotic--string 
models that do not contain any exotic massless states with fractional
electric charge. In this paper, we use this methodology to classify the
flipped $SU(5)$ free fermionic string models with respect to
some phenomenological criteria. 
For example, a question of interest is the existence
of viable three generation exophobic flipped $SU(5)$ vacua.
The observable sector of a heterotic--string flipped $SU(5)$
model is characterised by 15 integers
$\left(n_1,n_{\overline{1}},n_{5s},n_{\overline{5}s},n_{10},n_{\overline{10}},n_g,n_{{10H}},
n_{\overline{5}v},n_{5v},n_{{5h}},n_{1e},n_{\overline{1}e},n_{5e},n_{\overline{5}e}\right)$, which are given by:
\begin{align}
&n_{1}\,\,\,\,\,\,=\,\,\,\text{\# of } ({\textbf 1},{+\textstyle\frac{5}{2}}),\nn\\
&n_{\overline{1}}\,\,\,\,\,\,=\,\,\,\text{\# of } ({\textbf 1},{-\textstyle\frac{5}{2}}),\nn\\
&n_{5s}\,\,\,\,=\,\,\,\text{\# of } ({\textbf 5},{+\textstyle\frac{3}{2}}),\nn\\
&n_{\overline{5}s}\,\,\,\,=\,\,\,\text{\# of } ({\overline{\textbf 5}},{-\textstyle\frac{3}{2}}),\nn\\
&n_{10}\,\,\,\,=\,\,\,\text{\# of } ({\textbf {10}},{+\textstyle\frac{1}{2}}),\nn\\
&n_{\overline{10}}\,\,\,\,=\,\,\,\text{\# of } ({\overline{\textbf {10}}},{-\textstyle\frac{1}{2}}),\nn\\
&n_g\,\,\,\,\,\,=\,\,\,n_{10}-n_{\overline{10}}\,=\,\,\,n_{\overline{5}}-n_{5}=\,\,\,
  \text{\# of generations,}\nn\\
&n_{{10H}}=\,\,\,n_{10} + n_{\overline{10}}\,=\,\,\,\text{\# of non chiral heavy Higgs pairs,}\nn\\
&n_{\overline{5}v}\,\,\,\,=\,\,\,\text{\# of } ({\overline{\textbf 5}},{+1}),\nn\\
&n_{5v}\,\,\,\,=\,\,\,\text{\# of } ({\textbf 5},{-1}),\nn\\
&n_{{5h}}\,\,\,\,=\,\,\,n_{5v} + n_{\overline{5}v}=\,\,\,\text{\# of non chiral light Higgs pairs,}\nn\\
&n_{1e}\,\,\,\,\,=\,\,\,\text{\# of } ({\textbf 1 },
{-\tfrac{ 5}{ 4}})\text{ \,\, (exotic),}\nn\\
&n_{\overline{1}e}\,\,\,\,\,=\,\,\,\text{\# of } ({{\textbf 1}},
{+\tfrac{ 5}{ 4}})\text{ \,\, (exotic),}\nn\\
&n_{5e}\,\,\,\,\,=\,\,\,\text{\# of } ({\textbf 5},
{ -\tfrac{1}{4}} )\ \text{ \,\,(exotic),}\nn\\
&n_{\overline{5}e}\,\,\,\,\,=\,\,\,\text{\# of } ({\overline{\textbf 5}},
{+\tfrac{ 1}{ 4}}) \text{ \,\, (exotic).}\nn
\end{align}
These numbers above are all relevant for the classification of the string vacua. 
As noted in section \ref{chiralityoperators}, the $\alpha$ projection
dictates that $n_{\overline{1}}=n_{5s}$ and $n_{1}=n_{\overline{5}s}$.
Therefore, the counting of $n_{5}$ and $n_{\overline{5}}$ is sufficient for the number generations as shown above. Moreover, we note the distinction between the $\textbf5$ and $\overline{\textbf{5}}$ representations
that arise from the spinorial $\textbf{16}$ representation of $SO(10)$ decomposed under $SU(5) \times U(1)$, denoted by $n_{5s},n_{\overline{5}s}$
and the $\textbf5$ and $\overline{\textbf{5}}$ that arise from its vectorial $\textbf{10}$ representation, denoted by $n_{5v},n_{\overline{5}v}$.
While the former gives rise to the Standard Model up--type quark
electroweak singlet 
and lepton--doublet, 
the later accommodates the light electroweak Higgs doublets.
In the flipped $SU(5)$ models they are distinguished by their charges under
the $U(1)_5$ symmetry. 
Using the methodology outlined in section
\ref{projectors}, we obtained analytic formulas for all these quantities. In order to extract a
string spectrum from the phenomenologically viable models of the flipped $SU(5)$, we must have:
\begin{align}
&n_g=3~~~~~~~~~~~~\text{Three light chiral of generations,}\nn\\
&n_{{10H}}\ge1~~~~~~~~~\text{At least one heavy Higgs pair to 
break the $SU(5)\times U(1)$ symmetry,}\nn\\
&n_{{5}h}\ge1~~~~~~~~~~~\text{At least one pair of light Minimal SM Higgs doublets,}\nn\\
&n_{1e}=n_{\overline{1}e} \ge 0~~~
                          \text{Heavy mass can be generated for vector--like
exotics,}\nn\\
&n_{5e} = n_{\overline{5}e} \ge 0~~~
                          \text{Heavy mass can be generated for vector--like
exotics.}\nn\\ \nn
\end{align}
Here, we imposed the constraints 
$n_{\overline{5}h} = n_{5h}$, $n_{1e} = n_{\overline{1}e}$ and
$n_{5e} = n_{\overline{5}e}$, in order to sustain anomaly 
free flipped $SU(5)$ models.
\newpage
\subsection{Minimal exophilic models}
Compared to the case of the Pati--Salam classification
\cite {ACFKR2011}, which yielded 3 generation  models that are completely
free of massless exotic states, no such models were found 
in our scan of the flipped $SU(5)$ models.
We emphasise that this does not indicate that exophobic 
free fermionic flipped $SU(5)$ vacua do not exist, 
but merely that they do not exist in the space of vacua 
that we explored. Nevertheless, it does show that 
large spaces of vacua may not contain exophobic models, 
which is in line with related searches \cite{GatoRivera2011}.
A model with a minimal number of exotic states that we find in our
scan has
$n_g = 3$, $n_{\overline{5}s} = 3$,  $n_{5s} = 0$,
$n_{10} = 4$, $n_{\overline{10}}=1$, $n_{10H} = 1$,
$n_{5h} = 4$, $n_{1e} = 2$ and $n_{5e} = 0$.
We note that this minimal model still contains exotics which has 4 states.
The minimal model is then given by the following GGSO coefficients matrix:

\begin{equation} \label{BigMatrix}  (v_i|v_j)\ \ =\ \ \bordermatrix{
        & S&e_1&e_2&e_3&e_4&e_5&e_6&b_1&b_2&z_1&z_2&\alpha\cr
S  		& 1& 1& 1& 1& 1& 1& 1& 1& 1& 1& 1& 1\cr
e_1		& 1& 0& 0& 1& 1& 1& 0& 0& 0& 1& 1& 1\cr
e_2		& 1& 0& 0& 0& 0& 1& 0& 1& 0& 1& 1& 1\cr
e_3		& 1& 1& 0& 0& 0& 0& 0& 0& 0& 0& 1& 0\cr
e_4		& 1& 1& 0& 0& 0& 0& 0& 0& 0& 0& 1& 1\cr
e_5		& 1& 1& 1& 0& 0& 0& 1& 0& 1& 1& 1& 1\cr
e_6		& 1& 0& 0& 0& 0& 1& 0& 1& 1& 0& 1& 1\cr
b_1		& 0& 0& 1& 0& 0& 0& 1& 0& 0& 1& 1& 1/2\cr
b_2		& 0& 0& 0& 0& 0& 1& 1& 0& 1& 0& 0& 1/2\cr
z_1		& 1& 1& 1& 0& 0& 1& 0& 1& 0& 1& 1& 1\cr
z_2		& 1& 1& 1& 1& 1& 1& 1& 1& 0& 1& 0& -1/2\cr
\alpha	& 1& 1& 1& 0& 1& 1& 1& 0& 0& 0& 0& 1\cr
  }
\end{equation}
In section \ref{exoticstructure}
we elaborate further on the structure of the 
exotic states in the flipped $SU(5)$ models. 

\subsection{Classification}
\begin{figure}[ht!]
\centering
\includegraphics[width=140mm]{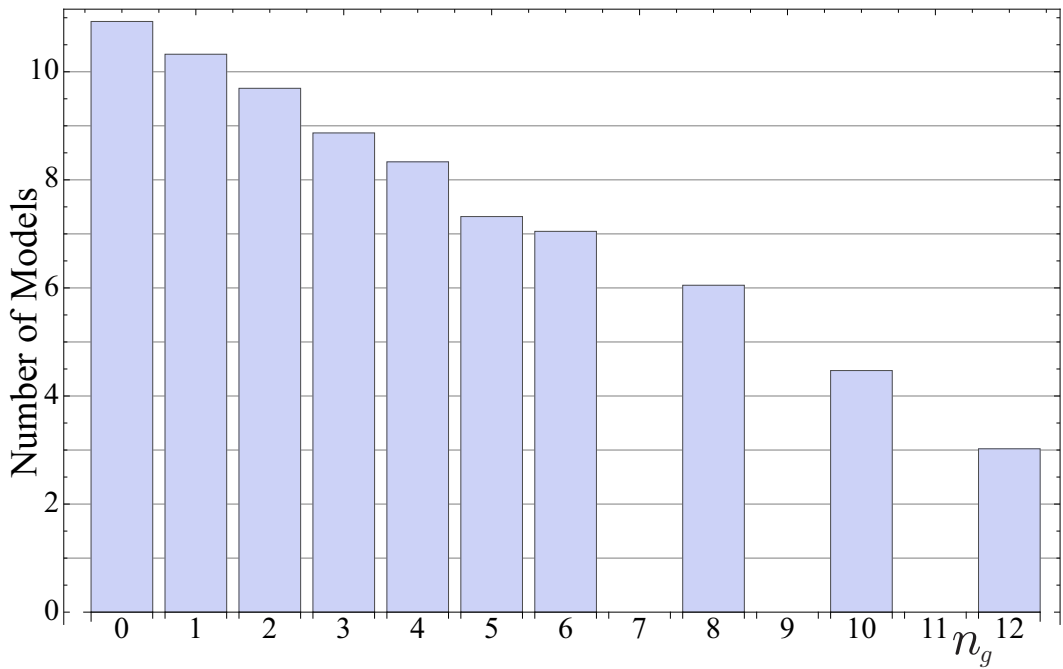}
\caption{\emph{Logarithm of the number of models 
against the number of generations ($n_g$)
in a random sample of $10^{12}$ flipped $SU(5)$ configurations.}}
\label{figure1}
\end{figure}

Next, we elaborate on the classification of the space of the free fermionic flipped $SU(5)$ 
string vacua by performing a statistical sampling, in a space
of $10^{12}$ models out of the $2^{44}$ possibilities. For 
this purpose, we developed two independent computer codes. 
One being a
FORTRAN95 computer program running on a single node of the Theoretical Physics
Division of University of Ioannina, HPC cluster.
The other being a JAVA code running on 10 nodes of the University of Liverpool, 
Department of Physics ULGQCD cluster that runs on AMD Opteron 6128 2GhZ CPUs.
Additionally,
assistance from five servers
of the University of Liverpool, Department of Mathematics 
were also used,
which totalled to 200 CPUs that ran for about 
2 weeks in order to scan $10^{12}$ random models.
Some of the results are presented in figures 1 - \ref{figure3} and table 1 - \ref{summarytable}.

In figure \ref{figure1}, the number of models against the
number of generations is displayed.
This is in agreement with the 
results of \cite{FKNR2004, FKR2007, ACFKR2011}, where
the number of models peaks for 0 generation and decreases as the number 
of generations increases. Also in this figure, we note 
the absence of models with
7, 9, 11 and greater than 12 generations. 
These result can be understood in light of the corresponding 
results in the $SO(10)$ classification \cite{FKR2007}. We recall that the 
$\alpha$ projection which breaks the $SO(10)$ symmetry 
to the flipped $SU(5)$, truncates the number of generations
by two. Examining the corresponding figure in the $SO(10)$ 
classification, we observe the absence of the 
models with double the number of generations, {\it i.e.} 
with 14, 18, 22 and more than 24 generations. We remark 
that this result is applicable to the case in 
which all gauge group enhancements are projected out,
as discussed in section \ref{gge}. Therefore, models with the 
excluded number of generations may occur when the hidden 
gauge group is enhanced. However, we compare figure \ref{figure1} to the corresponding figure in \cite{ACFKR2011} and notice the existence of models with 16 generations, which seems to contradict our argument. This apparent contradiction 
is resolved by noting that the $\alpha$ projection
in the Pati-Salam case projects out some of the gauge group enhancements, 
whereas from section \ref{gge} we note that this is 
not the case for the basis vector $\alpha_1$ we used here. Therefore, some of these models
descend from $SO(10)$ GUTs with enhanced gauge group, 
which do not arise in the case of the flipped $SU(5)$ models
studied here.

\begin{table}
\centering
\begin{tabular}{|c|r|r|r|r|}
\hline
$n_{5h}/n_{10H}$&0&1&2&3\\
\hline
0&281477&28518&0&0\\
\hline
1&3626622&275967&8197&651\\
\hline
2&630727&61910&2092&0\\
\hline
3&23924485&63774&5901&0\\
\hline
4&78959&67900&0&0\\
\hline
5&139642&12380&0&0\\
\hline
\end{tabular}
\caption{\label{num3gvshiggs} 
Number of three generation models as a function of the flipped $SU(5)$ breaking
Higgs pairs ($n_{10H}$) and SM breaking Higgs pairs ($n_{5h}$) in a random
sample of $10^{10}$ models. Models with $n_{10H}=0$ or $n_{5h}=0$ are not SM
compatible, the former because  $SU(5)$ cannot be broken to the SM via
the Higgs mechanism and the latter due to the lack of SM breaking Higgs
scalars. }
\end{table}

In table \ref{num3gvshiggs}, we display the number of three generation 
models against the number of pairs of light and heavy Higgs
representations appearing in the models, with the light and heavy pairs 
being $\textbf{5}+\overline{\textbf5}$ and $\textbf{10}+\overline{\textbf{10}}$ representations
of $SU(5)$, respectively. Clearly, the null cases are not
viable phenomenologically and the minimal cases are models with 
one pair of each. In models with a larger number of light Higgs pairs 
it may be easier to accommodate the Standard Model fermion mass
textures, whereas models with a larger number of heavy Higgs 
pairs may facilitate gauge coupling unification at the string scale
\cite{FSU51993, GCU1993}. 

As seen in section \ref{analysis4}, some of the exotic matter states
in the models transform in vector--like representations of the hidden
sector non--Abelian group factors.  
They carry fractional electric charge and must be sufficiently massive
or confined. These exotic states may nevertheless 
have interesting phenomenological implications. 
In table \ref{exotable}, we explore the structure of the 
exotic states arising in the models, which are 
labelled by four integers,
$({\bf n}_5^e, {\bf n}_1^e,{\bf n}_4^e, {\bf n}_{4^\prime}^e)$, 
where
${\bf n}_5^e= n_{5e}+n_{\overline{5}e}$
is the number of exotic states that transform as 
${\textbf 5}$ and ${\overline{\textbf 5}}$
of the observable $SU(5)$; 
${\bf n}_1^e= n_{1e}+n_{\overline{1}e}$
is the number of exotic states that transform as 
singlets of all non--Abelian group factors; 
${\bf n}_4^e= n_{4e}+n_{\overline{4}e}$
is the number of exotic states that transform as 
${\textbf 4}$ and ${\overline{\textbf 4}}$
of the hidden $SU(4)$;
${\bf n}_{4^\prime}^e= n_{{4^\prime}e}+n_{{\overline{4}^\prime}e}$
is the number of exotic states that transform as 
${\textbf 4}$ and ${\overline{\textbf 4}}$
of the hidden $SO(6)$ gauge group.

\begin{figure}[ht!]
\centering
\includegraphics[width=140mm]{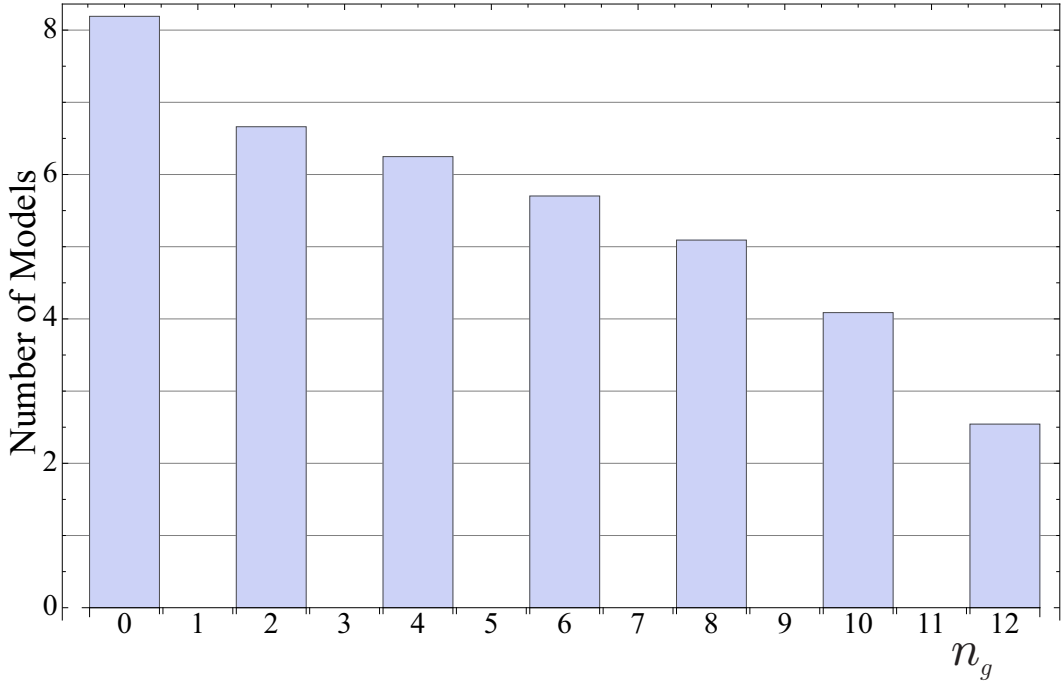}
\caption{\emph{Logarithm of the number of exophobic models against the number of generations ($n_g$)
in a random sample of $10^{12}$ flipped $SU(5)$ configurations.}}
\label{figure2}
\end{figure}

In figure \ref{figure2}, we display the number of exophobic models
against the number of generations. The striking feature in this figure
is the absence of models with three chiral generations. This is 
in contrast to the case of the Pati--Salam models that yielded 
numerous three generation exophobic models. Figure \ref{figure2} also reveals the absence of any exophobic models 
with 1, 3, 5, 7, 9, 11 and any exophobic models with more than
12 generation, whereas exophobic models arise for even number of 
generations, up to 12. As a result, exophobic models in this class 
arise in configurations with even number of generations but not in 
models with an odd number of generations.
We emphasize that these results hold in the space of models 
that we explore here and does indicate absence of three generation
exophobic flipped $SU(5)$ models. 
In figure \ref{figure3}, we display the number of three generation
models against the number of exotic multiplets. We note from the figure 
that the minimal number of exotic multiplets is 4.

\begin{figure}[ht!]
\centering
\includegraphics[width=140mm]{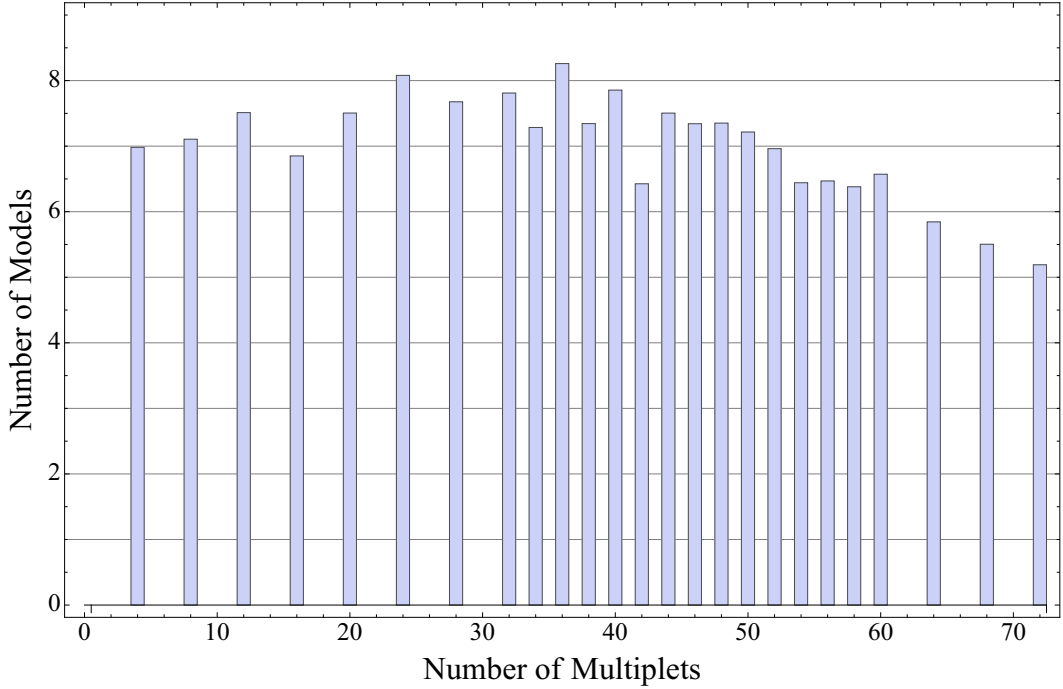}
\caption{\emph{Logarithm of the number 3 generation models against the number of exotic multiplets  ($n_e=n_{1e}+n_{\overline{1}e}+n_{5e}+n_{\overline{5}e}$)
in a random sample of $10^{12}$ flipped $SU(5)$ configurations.}}
\label{figure3}
\end{figure}

\subsection{The structure of the exotic states}\label{exoticstructure}

One of the main highlights of our classification method in the case of the 
Pati--Salam heterotic--string models has been the discovery of the 
exophobic heterotic--string models, in which all exotic states
are limited to the massive spectrum and do not appear among the 
massless states. As shown in figures \ref{figure2} and \ref{figure3},
in the class of $10^{12}$ flipped $SU(5)$ models that we analysed here, 
there are no exophobic 3 generation vacua with statistical a frequency larger than
$1:10^{12}$.
The structure of the exotic states arising in the models 
is analysed further in table \ref{exotable}.
All the models given in the table \ref{exotable}
contain three chiral generations of which at least one--pair is the
light Higgs states and at least one pair is the heavy Higgs states.
Thus, in all these models the gauge symmetry can be broken 
to the Standard Model in the effective field theory limit 
and contain all the fields required for viable Standard Model
phenomenology. 

From table \ref{exotable}, we note the occurrence of models 
in which all exotic states transform in representations 
of an hidden non--Abelian gauge group. In this case, the
exotic states are confined into integrally charged states 
and produce the so--called cryptons \cite{Cryptons1990}.
We further note from this table, the existence 
of models with equal number of $\textbf{4}$ and $\textbf{4}^\prime$ states. 
This suggests the possible existence of the free fermionic 
models that admit the race--track mechanism to stabilise
the vacuum expectation value of the dilaton field \cite{Racetrack1987}.
Moreover, table \ref{exotable} reveals interesting observations and 
directions for future research. The first eleven models in the 
table contain only states that transform in non--trivial 
representations of an hidden non--Abelian gauge group.
Thus, this class of models may give rise to the so--called 
crypton states that are confined into integrally charged
states. We see that there is an abundance 
of such models. There are also numerous models with small number
of crypton states that may remain asymptotically free and
therefore, confined at some scale. A well known example of a model
that gives rise only to crypton like states is given in \cite{RevampAEHN}. The table shows the existence of a large space 
of models with similar characteristics. One notable difference
between the vacua in this table and the one
of \cite{RevampAEHN}, is the fact that the model in 
\cite{RevampAEHN} uses asymmetric internal shifts, whereas
the models in this table only use symmetric 
internal shifts. The models in the six and twelfth
rows of the table, with $n_4=n_{4^\prime}=2$
are interesting to study for 
implementation of the racetrack mechanism \cite{Racetrack1987}.

Turning to the other types of exotic states.
The non--Abelian singlet states that are counted in the second 
column are fractionally charged and must decouple from the
light spectrum or be sufficiently diluted.
The fields counted in the first column transform
as $\bf{5}$ and $\overline{\bf{5}}$ of the observable $SU(5)$ and carry
$1/2$ of the hypercharge compared to the standard flipped
$SU(5)$ states. Such states do not arise in the flipped 
$SU(5)$ model studied in \cite{RevampAEHN}, but their 
colour triplet and electroweak doublet components
arise generically in the standard--like heterotic--string 
models \cite{SLM, GCU1993}. These fields may be instrumental 
as intermediate matter states to resolve the 
conflict between heterotic--string scale unification and
the low scale gauge coupling experimental data \cite{GCU1993}. 
The models appearing in the 13$^{\rm th}$ and 25$^{\rm rd}$
rows in table \ref{exotable}, are interesting examples of flipped 
$SU(5)$ models admitting such states. The models 
in the 25$^{\rm rd}$ row with $n_5=2$, $n_1=6$ , $n_4=2$
and $n_{4^\prime}=2$ may accommodate 
both the intermediate matter thresholds and
the racetrack mechanism, therefore be of
particular interest. 

\subsection{An illustrative example}

In this section, we analyse the model given in (\ref{BigMatrix2}) as 
an illustrative of the exotic spectrum appearing in the 
flipped $SU(5)$ models. 
The twisted sectors of the model given here
produce three chiral generations; 
one pair of heavy Higgs states; 
one pair of light Higgs representations. 
Therefore, this model may yield viable 
Standard Model phenomenology. 
\begin{equation}  \label{BigMatrix2}  (v_i|v_j)\ \ =\ \ \bordermatrix{
        & S&e_1&e_2&e_3&e_4&e_5&e_6&b_1&b_2&z_1&z_2&\alpha\cr
S  		& 1& 1& 1& 1& 1& 1& 1& 1& 1& 1& 1& 1\cr
e_1		& 1& 0& 0& 0& 0& 1& 1& 1& 0& 1& 0& 1\cr
e_2		& 1& 0& 0& 1& 1& 1& 1& 1& 0& 0& 0& 0\cr
e_3		& 1& 0& 1& 0& 0& 1& 1& 0& 1& 0& 1& 0\cr
e_4		& 1& 0& 1& 0& 0& 1& 1& 0& 1& 0& 1& 1\cr
e_5		& 1& 1& 1& 1& 1& 0& 0& 1& 1& 0& 0& 0\cr
e_6		& 1& 1& 1& 1& 1& 0& 0& 1& 1& 0& 0& 1\cr
b_1		& 0& 1& 1& 0& 0& 1& 1& 1& 0& 0& 0& -1/2\cr
b_2		& 0& 0& 0& 1& 1& 1& 1& 0& 1& 0& 0& 1/2\cr
z_1		& 1& 1& 0& 0& 0& 0& 0& 0& 0& 1& 1& 0\cr
z_2		& 1& 0& 0& 1& 1& 0& 0& 0& 0& 1& 0& 1/2\cr
\alpha	& 1& 1& 0& 0& 1& 0& 1& 1& 0& 1& 1& 1\cr
  }
\end{equation} 

Additionally, the model contains the following states 
that transform under the hidden $SU(4)$ gauge group: 
six non--exotic pairs of $(\textbf{4}+\overline{\textbf4})$; 
one non--exotic state transforming in the vectorial $\textbf6$ representation; 
one pair of exotic states transforming as $(\textbf{4}+\overline{\textbf4})$. 
The model contains the following states that transform
under the hidden $SU(4)^\prime$ gauge group: 
four non--exotic pairs of $(\textbf{4}+\overline{\textbf4})$; 
one non--exotic state transforming in the vectorial $\bf{6}$ representation; 
one pair of exotic states transforming as $(\textbf{4}+\overline{\textbf4})$. 
Thus, the $\beta$--functions of the $SU(4)$ and
$SU(4)^\prime$ hidden gauge groups are $\beta_4=-4$ 
and $\beta_{4^\prime}=-6$, respectively. Depending
on the mass scales for the hidden 
sector matter states, this model may therefore
provide workable example for implementing
the racetrack mechanism. The model also contains one
pair of exotic $(\textbf{5}+\overline{\textbf5})$ states of the observable 
flipped $SU(5)                                     $ group that can be used 
to mitigate the gauge coupling unification problem. 

\begin{table}
\begin{tabular}{|c|c|c|c|r||c|c|c|c|r||c|c|c|c|r|}
\hline
$n_{5}$&$n_1$&$n_4$&$n_4'$&\# &$n_{5}$&$n_1$&$n_4$&$n_4'$&\# &$n_{5}$&$n_1$&$n_4$&$n_4'$&\#\\
\hline
0 & 0 & 0 & 2 & 12627 & 2 & 10 & 3 & 3 & 9311 & 4 & 12 & 4 & 3 & 4889 \\
 0 & 0 & 0 & 4 & 3561 & 2 & 10 & 4 & 5 & 668 & 4 & 12 & 4 & 5 & 5720 \\
 0 & 0 & 0 & 6 & 1630 & 2 & 10 & 5 & 1 & 1614 & 4 & 12 & 4 & 6 & 965 \\
 0 & 0 & 0 & 8 & 187 & 2 & 10 & 5 & 2 & 4074 & 4 & 12 & 5 & 2 & 1479 \\
 0 & 0 & 2 & 0 & 16329 & 2 & 10 & 5 & 4 & 906 & 4 & 12 & 5 & 4 & 6105 \\
 0 & 0 & 2 & 2 & 18381 & 2 & 10 & 7 & 2 & 1745 & 4 & 12 & 6 & 4 & 608 \\
 0 & 0 & 2 & 4 & 2409 & 2 & 14 & 2 & 5 & 1474 & 4 & 12 & 8 & 0 & 153 \\
 0 & 0 & 4 & 0 & 6814 & 2 & 14 & 4 & 5 & 873 & 4 & 16 & 1 & 1 & 11395 \\
 0 & 0 & 4 & 2 & 3722 & 2 & 14 & 5 & 2 & 1412 & 5 & 7 & 2 & 4 & 1352 \\
 0 & 0 & 6 & 0 & 2338 & 2 & 14 & 5 & 4 & 1040 & 5 & 7 & 4 & 2 & 1323 \\
 0 & 0 & 8 & 0 & 356 & 2 & 18 & 1 & 1 & 2966 & 5 & 11 & 2 & 2 & 9462 \\
 0 & 8 & 2 & 2 & 1166 & 3 & 9 & 2 & 4 & 9505 & 5 & 11 & 2 & 5 & 2675 \\
 1 & 3 & 1 & 1 & 45575 & 3 & 9 & 3 & 3 & 2670 & 5 & 11 & 3 & 4 & 2491 \\
 1 & 3 & 2 & 8 & 4343 & 3 & 9 & 4 & 2 & 9949 & 5 & 11 & 4 & 3 & 2828 \\
 1 & 3 & 4 & 6 & 12465 & 3 & 13 & 2 & 2 & 9367 & 5 & 11 & 4 & 5 & 5164 \\
 1 & 3 & 6 & 4 & 12858 & 3 & 13 & 3 & 4 & 2562 & 5 & 11 & 5 & 2 & 2432 \\
 1 & 3 & 8 & 2 & 4287 & 3 & 13 & 4 & 3 & 2599 & 5 & 11 & 5 & 4 & 5074 \\
 1 & 11 & 2 & 4 & 1135 & 3 & 13 & 4 & 5 & 1909 & 5 & 15 & 2 & 5 & 4163 \\
 1 & 11 & 4 & 2 & 1336 & 3 & 13 & 5 & 4 & 1630 & 5 & 15 & 2 & 7 & 1069 \\
 2 & 6 & 0 & 0 & 13014 & 4 & 4 & 2 & 2 & 1231 & 5 & 15 & 4 & 5 & 1937 \\
 2 & 6 & 0 & 2 & 17622 & 4 & 8 & 1 & 5 & 1331 & 5 & 15 & 5 & 2 & 2605 \\
 2 & 6 & 0 & 4 & 6164 & 4 & 8 & 2 & 5 & 993 & 5 & 15 & 5 & 4 & 1170 \\
 2 & 6 & 0 & 6 & 3942 & 4 & 8 & 3 & 3 & 7915 & 5 & 15 & 7 & 2 & 670 \\
 2 & 6 & 2 & 0 & 14550 & 4 & 8 & 4 & 5 & 649 & 6 & 14 & 1 & 1 & 9970 \\
 2 & 6 & 2 & 2 & 25235 & 4 & 8 & 5 & 1 & 1443 & 6 & 18 & 0 & 2 & 2171 \\
 2 & 6 & 2 & 4 & 5864 & 4 & 8 & 5 & 2 & 1298 & 6 & 18 & 2 & 0 & 1499 \\
 2 & 6 & 3 & 5 & 10824 & 4 & 8 & 5 & 4 & 981 & 6 & 18 & 2 & 2 & 2272 \\
 2 & 6 & 4 & 0 & 5593 & 4 & 12 & 0 & 2 & 3255 & 6 & 18 & 2 & 4 & 799 \\
 2 & 6 & 4 & 2 & 4924 & 4 & 12 & 0 & 4 & 6986 & 6 & 18 & 4 & 2 & 490 \\
 2 & 6 & 4 & 4 & 2712 & 4 & 12 & 0 & 8 & 383 & 7 & 13 & 2 & 5 & 1311 \\
 2 & 6 & 4 & 6 & 1870 & 4 & 12 & 2 & 0 & 1795 & 7 & 13 & 2 & 7 & 758 \\
 2 & 6 & 5 & 3 & 10858 & 4 & 12 & 2 & 2 & 7064 & 7 & 13 & 5 & 2 & 849 \\
 2 & 6 & 6 & 0 & 2699 & 4 & 12 & 2 & 4 & 3139 & 7 & 13 & 7 & 2 & 428 \\
 2 & 6 & 6 & 4 & 2099 & 4 & 12 & 2 & 5 & 1269 & 8 & 12 & 1 & 1 & 2755 \\
 2 & 10 & 1 & 5 & 1522 & 4 & 12 & 3 & 4 & 5217 & 8 & 24 & 0 & 4 & 397 \\
 2 & 10 & 2 & 5 & 2794 & 4 & 12 & 4 & 0 & 3237 & 8 & 24 & 4 & 0 & 163 \\
 2 & 10 & 2 & 7 & 1199 & 4 & 12 & 4 & 2 & 2489 & - & - & - & - & -\\
 \hline
\end{tabular}
\caption{\label{exotable}
\emph{Number of three generation models with 
($n_{10H}\ge1,n_{5h}\ge1$) against fractional charge state multiplicities 
in a sample of $10^{10}$ randomly selected models. Here $n_5$ is the 
number of $\mathbf{5}+\mathbf{\overline{5}}$ $SU(5)$ pairs, $n_1$ is 
the number of fractional $SU(5)$ singlet pairs, $n_4$ is the number of 
$\mathbf{4}+\mathbf{\overline{4}}$ pairs transforming under hidden 
$SU(4)$ and $n'_4$ is the number of $\mathbf{4}+\mathbf{\overline{4}}$ 
pairs transforming under hidden $SU(4)'$.}}
\end{table}

\newpage

\section{Conclusion}
String theory continues to provide the only viable 
contemporary framework to explore the unification of
gravity with the gauge interactions. For this reason, models with three generations must be constructed phenomenologically. Whilst examples of such fully realistic models
may be a long way into the future, string theory
provides an abundance of these concrete quasi--realistic examples that 
can be explored as toy models on the way to achieving the
ultimate goal.

Here, we have continued to
develop the methodology for the classification
of the free fermionic heterotic--string models.
The free fermionic construction \cite{ABKKLT1987}, gave rise
to some of the most realistic string models constructed
to date \cite{RevampAEHN, ALR1990, SLM, LRS2001, ACFKR2011, CFR2011}.
These models correspond to ${\mathbb{Z}}_2 \times {\mathbb{Z}}_2$ orbifold
compactifications at special points in the moduli 
space with discrete Wilson lines \cite{Z2Z2Faraggi2002, Z2Z2Kounnas1997}.
The classification methodology was developed in
\cite{GKR1999} for the type II
superstrings and then adopted for the 
classification of free fermionic heterotic--string
models in \cite{FKNR2004, FKR2007, ACFKR2011}. In this 
method, the set of basis vectors are fixed incorporating 
all the possible symmetric ${\mathbb{Z}}_2$--shifts along the internal compactified
directions, in the form of the six basis vectors $e_i$ with adequate boundary conditions. 
The enumeration of the models is then obtained in terms 
of the GGSO projection coefficients and enables the scanning
of large number of models with the aid of computers. The initial classification in \cite{FKNR2004}
was with respect to chiral $\bf{16}$ and $\overline{\bf{16}}$ representations 
of an unbroken $SO(10)$ GUT group. Inclusion of
the $x$--map \cite{XMAP1993} in the methodology facilitated the classification with respect
to the vectorial $\bf{10}$ representations, which led to the discovery
of spinor--vector duality \cite{FKR2007, CFKR2009}. The 
classification methodology relies on writing
the GGSO projections in algebraic forms, which 
makes the enumeration of the massless spectrum
for a given configuration of GGSO phases straightforward.
This classification was also extended to 
models in which the $SO(10)$ symmetry was broken 
to the Pati--Salam subgroup leading to the discovery 
of exophobic string vacua \cite{ACFKR2011, CFR2011}. 

In this paper, we extended the development of the classification
methodology to the class of the free fermionic heterotic--string vacua
in which the $SO(10)$ symmetry is broken to the flipped $SU(5)$ 
subgroup. This case presents several complications with respect
to the previous ones. Whilst the earlier cases use 
only periodic and anti--periodic boundary conditions, 
the flipped $SU(5)$ class of vacua requires the use of
rational boundary conditions. Another, is the variation of the basis vectors that are used
to break the $SO(10)$ symmetry. With these modifications,
while adaptation of the algebraic expressions is readily available,
the computerised classification is substantially complicated. For this purpose, we developed two completely independent 
software routines, one based on JAVA and the second one being FORTRAN95
and all results presented in this paper are crossed checked 
using them. Our classification is limited 
to $1/2$ rational boundary conditions, which is the case 
in all the quasi--realistic free fermionic models to date.

{

\footnotesize
\begin{table}
\begin{tabular}{|c|l|r|c|c|r|}
\hline
&Constraints & \parbox[c]{2.5cm}{Total models in sample}& Probability
&\parbox[c]{3cm}{ Estimated number of models in class}\\
\hline
 & No Constraints & $1000000000000$ & $1$ &$1.76\times 10^{13}$ \\ \hline
(1)&{+ No Enhancements} & 762269298719 & $7.62\times 10^{-1}$ & $1.34\times
10^{13}$ \\  \hline
(2)&{+ Anomaly Free Flipped $SU(5)$} & 139544182312 & $1.40\times 10^{-1}$ & $2.45\times
10^{12}$ \\  \hline
(3)&{+ 3 Generations} & 738045321 & $7.38 \times 10^{-4}$ & $1.30\times
10^{10}$ \\  \hline
(4a)&{+ SM Light Higgs} & 706396035 & $7.06 \times 10^{-4}$ & $1.24\times
10^{10}$ \\\hline
(4b)&{+ Flipped $SU(5)$ Heavy Higgs} & 46470138 & $4.65 \times 10^{-5}$ & $8.18\times
10^{8}$ \\\hline
(5)&{+ SM Light Higgs}& 43624911 & $4.36 \times 10^{-5}$ & $7.67\times
10^8$ \\ &{+ \& Heavy Higgs}&&&  \\\hline
(6a)&{\parbox[c]{4cm}{+ Minimal Flipped $SU(5)$ Heavy Higgs}} & 42310396 & $4.23 \times 10^{-5}$ & $7.44\times
10^8$ \\\hline
(6b)&{+ Minimal SM Light Higgs } & 25333216 & $2.53 \times 10^{-5}$ & $4.46\times
10^8$ \\\hline
(7)&\parbox[c]{4cm}{+ Minimal Flipped $SU(5)$ Heavy Higgs} & 24636896 & $2.46 \times 10^{-5}$ & $4.33\times
10^8$\\ &{+ \& Minimal SM Light Higgs}&&&  \\\hline
(8)&{+ Minimal Exotic States} & 1218684 & $1.22 \times 10^{-6}$ & $2.14\times
10^7$ \\\hline
\end{tabular}
\caption{\label{summarytable} \emph{Statistics for the flipped $SU(5)$ models with respect to phenomenological constraints. Here we note that the results of 4a and 4b have no effect on each other and this also holds for 6a and 6b.}}
\end{table}

}

In table \ref{summarytable}, we tabulate the number of models 
with sequential imposition of phenomenological constraints.
The total number of models in the sample is $10^{12}$.
We first impose that there is no enhancement of the four dimensional 
gauge symmetry and there are approximately 76.2\% of the models that satisfy this criteria. 
Next we impose, that the flipped $SU(5)$ models are anomaly free with respect to the $U(1)_5$ group factor and about 14\% of the total models satisfy this criterion.
A further reduction, by three orders of magnitude, results from the restriction to the three chiral generations.
Next, imposing the existence of both the heavy and light Higgs states to break the
flipped $SU(5)$ gauge symmetry to the Standard Model gauge
group and the electroweak breaking respectively, leads to a further reduction i.e. one order of magnitude. Finally, imposing the minimal number of massless exotic states results in the
reduction of the number of models
by a further order of magnitude. Therefore, the number of string vacua in the space of models scanned, reduces from $10^{12}$ to $10^6$ which satisfy all the constraints that were imposed. This leaves a substantial number to accommodate
further phenomenological constraints.
In conclusion, we comment on the results obtained by using $\alpha_2$ 
and $\alpha_3$ in (\ref{alphas}) to break the $SO(10)$ symmetry.
In both cases a JAVA code was used to classify the models. The results are not that substantially different 
compared to the classification with $\alpha_1$ and we also do not
find any three generation exophobic vacua in these cases. 
\newpage
\section{Acknowledgements}

\normalsize
We would like to thank Laura Bernard and Ivan Glasser
for collaboration in the initial stages of this project. 
JR would like to thank the University of Liverpool and CERN, AEF would like to thank CERN, the University of Ioannina
and Oxford University for their hospitality, and HS would like to thank Johar Ashfaque for the many useful discussions.
AEF is supported in part by STFC under contract ST/J000493/1.
JR's work has been supported in part by
the ITN network ``UNILHC'' (PITN-GA-2009-237920). HS is supported by the STFC studentship award.
This research has been co-financed by the European Union (European 
Social Fund-ESF) and Greek national funds through the 
Operational Program "Education and Lifelong Learning", of the National 
Strategic Reference Framework (NSRF) (Thales Research Funding Program) investing in knowledge society through the European Social Fund.

\bigskip

\appendix

\section{ Projectors and matrix formalism}\label{appendixa}

The algebraic expressions corresponding to states in the string spectrum are given by: $\B{4,5,6}$ from (\ref{nonchiralvectorials}), which produce light Higgs and Hidden vectorial states; $\B{7,8,9}$ and $\B{10,11,12}$ given in (\ref{hidspin2}) and (\ref{hidspin3}) respectively, which produce spinorial hidden matter states; $\B{13,14,15}$ and $\B{16,17,18}$ given in (\ref{exoticspin1}) and (\ref{exoticspin2}) respectively, which produce spinorial exotic states; $\B{19,20,21}$ and $\B{22,23,24}$ given in (\ref{vecto}) and (\ref{bz1alpha}) respectively, which produce vectorial exotic states. We now enumerate these projectors with their corresponding algebraic expressions and matrix equations as follows.

\normalsize
\subsection{Vectorial representations}

The sectors in (\ref{nonchiralvectorials})
produce vectorial states in the observable and hidden sector. 
These sectors are obtained from the combinations

\begin{align*}
\B{4,5,6}&=\B{1,2,3}+z_1+2\alpha
\end{align*}

The following is a list of the states produced in these sectors
and the projectors that act on them: 
\begin{itemize}
\newpage
\item
\textbf{States 
$\{\overline\eta^{1,2,3}\}|R\rangle$, $\{\overline\eta^{*1,2,3}\}|R\rangle$, 
$\{\overline\psi^{1,...,5}\}|R\rangle$ and $\{\overline\psi^{*1,...,5}\}|R\rangle$}

This gives rise to the states that transform under the 
$SU(5) \times U(1)_5$ or $U(1)_{1/2/3}$ gauge group.
The projectors are given by:

\begin{align*}
{P_{pqrs}^{(4)(\overline\eta^{1},\overline\psi^{1,...,5})}} = 
\frac{1}{16} &\left( 1-C \binom {e_{1}} {\B{4}}\right) \cdot 
\left( 1-C \binom {e_{2}} {\B{4}}\right) \nonumber \\
\cdot &\left( 1-C \binom {z_1} {\B{4}}\right) \cdot 
\left( 1- C \binom {z_2} {\B{4}}\right) \nonumber \\
{P_{pqrs}^{(5)(\overline\eta^{2},\overline\psi^{1,...,5})}} = \frac{1}{16} &
\left( 1-C \binom {e_{3}} {\B{5}}\right) \cdot 
\left( 1-C \binom {e_{4}} {\B{5}}\right) \nonumber \\
\cdot &\left( 1-C \binom {z_1} {\B{5}}\right) \cdot 
\left( 1- C \binom {z_2} {\B{5}}\right)  \\
{P_{pqrs}^{(6)(\overline\eta^{3},\overline\psi^{1,...,5})}} = 
\frac{1}{16} &\left( 1-C \binom {e_{5}} {\B{6}}\right) \cdot 
\left( 1-C \binom {e_{6}} {\B{6}}\right) \nonumber \\
\cdot &\left( 1-C \binom {z_1} {\B{6}}\right) \cdot 
\left( 1- C \binom {z_2} {\B{6}}\right) \nonumber
\end{align*}

The corresponding matrix equations are given as:

\begin{align*}
\begin{pmatrix} (e_1|e_3)&(e_1|e_4)&(e_1|e_5)&(e_1|e_6)\\
(e_2|e_3)&(e_2|e_4)&(e_2|e_5)&(e_2|e_6)\\
(z_1|e_3)&(z_1|e_4)&(z_1|e_5)&(z_1|e_6)\\
(z_2|e_3)&(z_2|e_4)&(z_2|e_5)&(z_2|e_6)\\
\end{pmatrix}
\begin{pmatrix} p \\ q \\ r \\ s \end{pmatrix} &=
\begin{pmatrix} (e_1|b_1+z_1)\\
(e_2|b_1+z_1)\\
(z_1|b_1+z_1)\\
(z_2|b_1+z_1) + 1\\
\end{pmatrix} \nonumber
\\[0.3cm]
\begin{pmatrix} (e_3|e_1)&(e_3|e_2)&(e_3|e_5)&(e_3|e_6)\\
(e_4|e_1)&(e_4|e_2)&(e_4|e_5)&(e_4|e_6)\\
(z_1|e_1)&(z_1|e_2)&(z_1|e_5)&(z_1|e_6)\\
(z_2|e_1)&(z_2|e_2)&(z_2|e_5)&(z_2|e_6)\\
\end{pmatrix}
\begin{pmatrix} p \\ q \\ r \\ s \end{pmatrix} &=
\begin{pmatrix} (e_3|b_2+z_1)\\
(e_4|b_2+z_1)\\
(z_1|b_2+z_1)\\
(z_2|b_2+z_1) + 1\\
\end{pmatrix}
\\[0.3cm]
\begin{pmatrix} (e_5|e_1)&(e_5|e_2)&(e_5|e_3)&(e_5|e_4)\\
(e_6|e_1)&(e_6|e_2)&(e_6|e_3)&(e_6|e_4)\\
(z_1|e_1)&(z_1|e_2)&(z_1|e_3)&(z_1|e_4)\\
(z_2|e_1)&(z_2|e_2)&(z_2|e_3)&(z_2|e_4)\\
\end{pmatrix}
\begin{pmatrix} p \\ q \\ r \\ s \end{pmatrix} &=
\begin{pmatrix} (e_5|b_1+b_2)\\
(e_6|b_1+b_2)\\
(z_1|b_1+b_2)\\
(z_2|b_1+b_2)\\
\end{pmatrix} \nonumber
\end{align*}

\newpage
\item
\textbf{States $\{\overline\phi^{1,...,4}\}|R\rangle$ and $\{\overline\phi^{*1,...,4}\}|R\rangle$}

These states transform under the 
$SU(4) \times U(1)_4$ hidden gauge group. 
The projectors are given by:

\begin{align*}
{P_{pqrs}^{(4)(\overline\phi^{1,...,4})}} = \frac{1}{16} 
&\left( 1-C \binom {e_{1}} {\B{4}}\right) \cdot 
\left( 1-C \binom {e_{2}} {\B{4}}\right) \nonumber \\
\cdot &\left( 1+C \binom {z_1} {\B{4}}\right) \cdot 
\left( 1- C \binom {z_2} {\B{4}}\right) \nonumber \\
{P_{pqrs}^{(5)(\overline\phi^{1,...,4})}} = \frac{1}{16} 
&\left( 1-C \binom {e_{3}} {\B{5}}\right) \cdot 
\left( 1-C \binom {e_{4}} {\B{5}}\right) \nonumber \\
\cdot &\left( 1+C \binom {z_1} {\B{5}}\right) \cdot 
\left( 1- C \binom {z_2} {\B{5}}\right)  \\
{P_{pqrs}^{(6)(\overline\phi^{1,...,4})}} = \frac{1}{16} 
&\left( 1-C \binom {e_{5}} {\B{6}}\right) \cdot 
\left( 1-C \binom {e_{6}} {\B{6}}\right) \nonumber \\
\cdot &\left( 1+C \binom {z_1} {\B{6}}\right) \cdot 
\left( 1- C \binom {z_2} {\B{6}}\right) \nonumber
\end{align*}

The corresponding matrix equations are given as:

\begin{align*}
\begin{pmatrix} (e_1|e_3)&(e_1|e_4)&(e_1|e_5)&(e_1|e_6)\\
(e_2|e_3)&(e_2|e_4)&(e_2|e_5)&(e_2|e_6)\\
(z_1|e_3)&(z_1|e_4)&(z_1|e_5)&(z_1|e_6)\\
(z_2|e_3)&(z_2|e_4)&(z_2|e_5)&(z_2|e_6)\\
\end{pmatrix}
\begin{pmatrix} p \\ q \\ r \\ s \end{pmatrix} &=
\begin{pmatrix} (e_1|b_1+z_1)\\
(e_2|b_1+z_1)\\
(z_1|b_1+z_1) + 1\\
(z_2|b_1+z_1) + 1\\
\end{pmatrix} \nonumber
\\[0.3cm]
\begin{pmatrix} (e_3|e_1)&(e_3|e_2)&(e_3|e_5)&(e_3|e_6)\\
(e_4|e_1)&(e_4|e_2)&(e_4|e_5)&(e_4|e_6)\\
(z_1|e_1)&(z_1|e_2)&(z_1|e_5)&(z_1|e_6)\\
(z_2|e_1)&(z_2|e_2)&(z_2|e_5)&(z_2|e_6)\\
\end{pmatrix}
\begin{pmatrix} p \\ q \\ r \\ s \end{pmatrix} &=
\begin{pmatrix} (e_3|b_2+z_1)\\
(e_4|b_2+z_1)\\
(z_1|b_2+z_1) + 1\\
(z_2|b_2+z_1) + 1\\
\end{pmatrix}
\\[0.3cm]
\begin{pmatrix} (e_5|e_1)&(e_5|e_2)&(e_5|e_3)&(e_5|e_4)\\
(e_6|e_1)&(e_6|e_2)&(e_6|e_3)&(e_6|e_4)\\
(z_1|e_1)&(z_1|e_2)&(z_1|e_3)&(z_1|e_4)\\
(z_2|e_1)&(z_2|e_2)&(z_2|e_3)&(z_2|e_4)\\
\end{pmatrix}
\begin{pmatrix} p \\ q \\ r \\ s \end{pmatrix} &=
\begin{pmatrix} (e_5|b_1+b_2)\\
(e_6|b_1+b_2)\\
(z_1|b_1+b_2) + 1\\
(z_2|b_1+b_2)\\
\end{pmatrix} \nonumber
\end{align*}

\newpage
\item
\textbf{State $\{\overline\phi^{5,...,8}\}|R\rangle$ and $\{\overline\phi^{*5,...,8}\}|R\rangle$}

These states transform under the $U(1)_{hid}$ or $SO(6)$ 
gauge groups. The projectors on these states are given by:

\begin{align*}
{P_{pqrs}^{(4)(\overline\phi^{5,...,8})}} = \frac{1}{16} 
&\left( 1-C \binom {e_{1}} {\B{4}}\right) \cdot 
\left( 1-C \binom {e_{2}} {\B{4}}\right) \nonumber \\
\cdot &\left( 1-C \binom {z_1} {\B{4}}\right) \cdot 
\left( 1 + C \binom {z_2} {\B{4}}\right) \nonumber \\
{P_{pqrs}^{(5)(\overline\phi^{5,...,8})}} = 
\frac{1}{16} &\left( 1-C \binom {e_{3}} {\B{5}}\right) \cdot 
\left( 1-C \binom {e_{4}} {\B{5}}\right) \nonumber \\
\cdot &\left( 1-C \binom {z_1} {\B{5}}\right) \cdot 
\left( 1 + C \binom {z_2} {\B{5}}\right)  \\
{P_{pqrs}^{(6)(\overline\phi^{5,...,8})}} = \frac{1}{16} 
&\left( 1-C \binom {e_{5}} {\B{6}}\right) \cdot \left( 1-C 
\binom {e_{6}} {\B{6}}\right) \nonumber \\
\cdot &\left( 1-C \binom {z_1} {\B{6}}\right) \cdot 
\left( 1 + C \binom {z_2} {\B{6}}\right) \nonumber
\end{align*}

The corresponding matrix equations are given as:

\begin{align*}
\begin{pmatrix} (e_1|e_3)&(e_1|e_4)&(e_1|e_5)&(e_1|e_6)\\
(e_2|e_3)&(e_2|e_4)&(e_2|e_5)&(e_2|e_6)\\
(z_1|e_3)&(z_1|e_4)&(z_1|e_5)&(z_1|e_6)\\
(z_2|e_3)&(z_2|e_4)&(z_2|e_5)&(z_2|e_6)\\
\end{pmatrix}
\begin{pmatrix} p \\ q \\ r \\ s \end{pmatrix} &=
\begin{pmatrix} (e_1|b_1+z_1)\\
(e_2|b_1+z_1)\\
(z_1|b_1+z_1)\\
(z_2|b_1+z_1)\\
\end{pmatrix} \nonumber
\\[0.3cm]
\begin{pmatrix} (e_3|e_1)&(e_3|e_2)&(e_3|e_5)&(e_3|e_6)\\
(e_4|e_1)&(e_4|e_2)&(e_4|e_5)&(e_4|e_6)\\
(z_1|e_1)&(z_1|e_2)&(z_1|e_5)&(z_1|e_6)\\
(z_2|e_1)&(z_2|e_2)&(z_2|e_5)&(z_2|e_6)\\
\end{pmatrix}
\begin{pmatrix} p \\ q \\ r \\ s \end{pmatrix} &=
\begin{pmatrix} (e_3|b_2+z_1)\\
(e_4|b_2+z_1)\\
(z_1|b_2+z_1)\\
(z_2|b_2+z_1)\\
\end{pmatrix}
\\[0.3cm]
\begin{pmatrix} (e_5|e_1)&(e_5|e_2)&(e_5|e_3)&(e_5|e_4)\\
(e_6|e_1)&(e_6|e_2)&(e_6|e_3)&(e_6|e_4)\\
(z_1|e_1)&(z_1|e_2)&(z_1|e_3)&(z_1|e_4)\\
(z_2|e_1)&(z_2|e_2)&(z_2|e_3)&(z_2|e_4)\\
\end{pmatrix}
\begin{pmatrix} p \\ q \\ r \\ s \end{pmatrix} &=
\begin{pmatrix} (e_5|b_1+b_2)\\
(e_6|b_1+b_2)\\
(z_1|b_1+b_2)\\
(z_2|b_1+b_2) + 1\\
\end{pmatrix} \nonumber
\end{align*}

\end{itemize}

\newpage
\subsection{
Hidden sector representations
}

The sectors $\B{1,2,3}+2\alpha$ and $\B{1,2,3}+z_1+z_2+2\alpha$
give rise to non--exotic states that transform under the hidden gauge 
group. The states in these sectors descend from the ${\bf 16}$ 
vectorial representation of the hidden $SO(16)$ gauge group, 
decomposed under the final unbroken hidden sector gauge group. 
The sectors 
\begin{align*}
\B{7,8,9}&=\B{1,2,3}+2\alpha
\end{align*}
produce states that transform 
under the $SU(4) \times U(1)_4$ hidden gauge group. 
The projectors on states arising in these sectors are given by:
\begin{align*}
\PP{7} &= \frac{1}{8} \left( 1-C \binom {e_{1}} {\B{7}}\right)\cdot \left( 1-C \binom {e_{2}} {\B{7}}\right) \cdot \left( 1-C \binom {z_2} {\B{7}}\right) \nonumber \\
\PP{8} &= \frac{1}{8} \left( 1-C \binom {e_{3}} {\B{8}}\right)\cdot \left( 1-C \binom {e_{4}} {\B{8}}\right) \cdot \left( 1-C \binom {z_2} {\B{8}}\right) \\
\PP{9} &= \frac{1}{8} \left( 1-C \binom {e_{5}} {\B{9}}\right)\cdot \left( 1-C \binom {e_{6}} {\B{9}}\right) \cdot \left( 1-C \binom {z_2} {\B{9}}\right) \nonumber
\end{align*}
The corresponding matrix equations are given as:

\begin{align*}
\begin{pmatrix} (e_1|e_3)&(e_1|e_4)&(e_1|e_5)&(e_1|e_6)\\
(e_2|e_3)&(e_2|e_4)&(e_2|e_5)&(e_2|e_6)\\
(z_2|e_3)&(z_2|e_4)&(z_2|e_5)&(z_2|e_6) \end{pmatrix}
\begin{pmatrix} p\\q\\r\\s\end{pmatrix} &=
\begin{pmatrix} (e_1|b_1)\\
(e_2|b_1)\\
(z_2|b_1) + 1
\end{pmatrix} \nonumber
\\[0.3cm]
\begin{pmatrix} (e_3|e_1)&(e_3|e_2)&(e_3|e_5)&(e_3|e_6)\\
(e_4|e_1)&(e_4|e_2)&(e_4|e_5)&(e_4|e_6)\\
(z_2|e_1)&(z_2|e_2)&(z_2|e_5)&(z_2|e_6) \end{pmatrix}
\begin{pmatrix} p\\q\\r\\s\end{pmatrix} &=
\begin{pmatrix} (e_3|b_2)\\
(e_4|b_2)\\
(z_2|b_2) + 1
\end{pmatrix}
\\[0.3cm]
\begin{pmatrix} (e_5|e_1)&(e_5|e_2)&(e_5|e_3)&(e_5|e_4)\\
(e_6|e_1)&(e_6|e_2)&(e_6|e_3)&(e_6|e_4)\\
(z_2|e_1)&(z_2|e_2)&(z_2|e_3)&(z_2|e_4) \end{pmatrix}
\begin{pmatrix} p\\q\\r\\s\end{pmatrix} &=
\begin{pmatrix} (e_5|b_1+b_2+z_1)\\
(e_6|b_1+b_2+z_1)\\
(z_2|b_1+b_2+z_1)
\end{pmatrix} \nonumber
\end{align*}

\newpage
The sectors 
\begin{align*}
\B{10,11,12}&=\B{1,2,3} + z_1 + z_2 + 2\alpha
\end{align*}
give rise to states that transform under the hidden 
$SO(6)$ gauge group. 
The projectors acting on these states are given by:
\begin{align*}
\PP{10} = \frac{1}{8} \left( 1-C \binom {e_{1}} 
{\B{10}}\right) \cdot \left( 1-C \binom {e_{2}} 
{\B{10}}\right) \cdot \left( 1-C \binom {z_1} {\B{10}}\right) \nonumber \\
\PP{11} = \frac{1}{8} \left( 1-C \binom {e_{3}} 
{\B{11}}\right) \cdot \left( 1-C \binom {e_{4}} 
{\B{11}}\right) \cdot \left( 1-C \binom {z_1} {\B{11}}\right) \\
\PP{12} = \frac{1}{8} \left( 1-C \binom {e_{5}} 
{\B{12}}\right) \cdot \left( 1-C \binom {e_{6}} 
{\B{12}}\right) \cdot \left( 1-C \binom {z_1} {\B{12}}\right) \nonumber
\end{align*}
The corresponding matrix equations are given as:

\begin{align*}
\begin{pmatrix} (e_1|e_3)&(e_1|e_4)&(e_1|e_5)&(e_1|e_6)\\
(e_2|e_3)&(e_2|e_4)&(e_2|e_5)&(e_2|e_6)\\
(z_1|e_3)&(z_1|e_4)&(z_1|e_5)&(z_1|e_6) \end{pmatrix}
\begin{pmatrix} p\\q\\r\\s\end{pmatrix} &=
\begin{pmatrix} (e_1|b_1+z_1+z_2)\\
(e_2|b_1+z_1+z_2)\\
(z_1|b_1+z_1+z_2)
\end{pmatrix} \nonumber
\\[0.3cm]
\begin{pmatrix} (e_3|e_1)&(e_3|e_2)&(e_3|e_5)&(e_3|e_6)\\
(e_4|e_1)&(e_4|e_2)&(e_4|e_5)&(e_4|e_6)\\
(z_1|e_1)&(z_1|e_2)&(z_1|e_5)&(z_1|e_6) \end{pmatrix}
\begin{pmatrix} p\\q\\r\\s\end{pmatrix} &=
\begin{pmatrix} (e_3|b_2+z_1+z_2)\\
(e_4|b_2+z_1+z_2)\\
(z_1|b_2+z_1+z_2)
\end{pmatrix}
\\[0.3cm]
\begin{pmatrix} (e_5|e_1)&(e_5|e_2)&(e_5|e_3)&(e_5|e_4)\\
(e_6|e_1)&(e_6|e_2)&(e_6|e_3)&(e_6|e_4)\\
(z_1|e_1)&(z_1|e_2)&(z_1|e_3)&(z_1|e_4) \end{pmatrix}
\begin{pmatrix} p\\q\\r\\s\end{pmatrix} &=
\begin{pmatrix} (e_5|b_1+b_2+z_2)\\
(e_6|b_1+b_2+z_2)\\
(z_1|b_1+b_2+z_2)
\end{pmatrix} \nonumber
\end{align*}

\subsection{Exotics}
The exotic states are obtained from sectors containing the $SO(10)$ breaking vector 
$\alpha$. As mentioned in section \ref{analysis4}, the sectors 
that give rise to exotic states are classified according
to their vacuum in the right--moving sector. For a given sector $\xi$ 
with $\xi_R\cdot\xi_R=6$ a right--moving oscillator of a world--sheet 
fermion with $\pm1/4$ boundary conditions acting on the vacuum
is needed to obtain a massless state. Sectors with $\xi_R\cdot\xi_R=8$
produce massless states without an oscillator. The first type 
of sectors can therefore produce states that transform 
as $\bf{5}$ and $\overline{\bf{5}}$, as well as states that transform as singlets
under the observable $SU(5)$ gauge group.
The second type of sectors gives rise to states that transform
as singlets of the observable $SU(5)$ gauge symmetry.
All the exotic states transform in standard representations
under the observable $SU(5)$ 
gauge group (including singlets)
but carry exotic charge under the observable $U(1)_5$ gauge group.
The sectors 
\begin{align*}
\B{13,14,15}&=\B{1,2,3} + z_2 + \alpha
\end{align*}
produce states that 
transform under the $\bf{4}$ and $\overline{\bf{4}}$ of the $SO(6)$ hidden gauge group.
The projectors acting on these states are given by:

\begin{align*}
\PP{13} = \frac{1}{16} &\left( 1-C \binom {e_{1}} {\B{13}} \right) \cdot 
\left( 1-C \binom {e_{2}} {\B{13}} \right) \nonumber \\
& \left( 1+C \binom {z_1} {\B{13}} \right) \cdot 
\left( 1-C \binom {\alpha} {\B{13}} \right) \nonumber \\
\PP{14} = \frac{1}{16} &\left( 1-C \binom {e_{3}} {\B{14}} \right) \cdot 
\left( 1-C \binom {e_{4}} {\B{14}} \right) \nonumber \\
& \left( 1+C \binom {z_1} {\B{14}} \right) \cdot 
\left( 1-C \binom {\alpha} {\B{14}} \right) \\
\PP{15} = \frac{1}{16} &\left( 1-C \binom {e_{5}} {\B{15}} \right) \cdot 
\left( 1-C \binom {e_{6}} {\B{15}} \right) \nonumber \\
& \left( 1+C \binom {z_1} {\B{15}} \right) \cdot 
\left( 1-C \binom {\alpha} {\B{15}} \right) \nonumber
\end{align*}
The corresponding matrix equations are given as:

\begin{align*}
\begin{pmatrix} (e_1|e_3)&(e_1|e_4)&(e_1|e_5)&(e_1|e_6)\\
(e_2|e_3)&(e_2|e_4)&(e_2|e_5)&(e_2|e_6)\\
(z_1|e_3)&(z_1|e_4)&(z_1|e_5)&(z_1|e_6)\\
(\alpha|e_3)&(\alpha|e_4)&(\alpha|e_5)&(\alpha|e_6)
\end{pmatrix}
\begin{pmatrix} p\\q\\r\\s\end{pmatrix} &=
\begin{pmatrix} (e_1|b_1 + z_2 + \alpha)\\
(e_2|b_1 + z_2 + \alpha)\\
(z_1|b_1 + z_2 + \alpha) + 1\\
(\alpha|b_1 + z_2 + \alpha)\\
\end{pmatrix} \nonumber
\\[0.3cm]
\begin{pmatrix} (e_3|e_1)&(e_3|e_2)&(e_3|e_5)&(e_3|e_6)\\
(e_4|e_1)&(e_4|e_2)&(e_4|e_5)&(e_4|e_6)\\
(z_1|e_1)&(z_1|e_2)&(z_1|e_5)&(z_1|e_6)\\
(\alpha|e_1)&(\alpha|e_2)&(\alpha|e_5)&(\alpha|e_6)
\end{pmatrix}
\begin{pmatrix} p\\q\\r\\s\end{pmatrix} &=
\begin{pmatrix} (e_3|b_2 + z_2 + \alpha)\\
(e_4|b_2 + z_2 + \alpha)\\
(z_1|b_2 + z_2 + \alpha) + 1\\
(\alpha|b_2 + z_2 + \alpha)\\
\end{pmatrix}
\\[0.3cm]
\begin{pmatrix} (e_5|e_1)&(e_5|e_2)&(e_5|e_3)&(e_5|e_4)\\
(e_6|e_1)&(e_6|e_2)&(e_6|e_3)&(e_6|e_4)\\
(z_1|e_1)&(z_1|e_2)&(z_1|e_3)&(z_1|e_4)\\
(\alpha|e_1)&(\alpha|e_2)&(\alpha|e_3)&(\alpha|e_4)
\end{pmatrix}
\begin{pmatrix} p\\q\\r\\s\end{pmatrix} &=
\begin{pmatrix} (e_5|b_3 + z_2 + \alpha)\\
(e_6|b_3 + z_2 + \alpha)\\
(z_1|b_3 + z_2 + \alpha) + 1\\
(\alpha|b_3 + z_2 + \alpha)\\
\end{pmatrix} \nonumber
\end{align*}

\newpage
Similar exotic states are produced from the sectors:
\begin{align*}
\B{16,17,18}&=\B{1,2,3} + z_1 + z_2 + \alpha
\end{align*}
The projectors acting on these states are given by:

\begin{align*}
\PP{16} = \frac{1}{16} &\left( 1-C \binom {e_{1}} {\B{16}} \right) \cdot 
\left( 1-C \binom {e_{2}} {\B{16}} \right) \nonumber \\
& \left( 1+C \binom {z_1} {\B{16}} \right) \cdot 
\left( 1+C \binom {\alpha} {\B{16}} \right) \nonumber \\
\PP{17} = \frac{1}{16} &\left( 1-C \binom {e_{3}} {\B{17}} \right) \cdot 
\left( 1-C \binom {e_{4}} {\B{17}} \right) \nonumber \\
& \left( 1+C \binom {z_1} {\B{17}} \right) \cdot 
\left( 1+C \binom {\alpha} {\B{17}} \right) \\
\PP{18} = \frac{1}{16} &\left( 1-C \binom {e_{5}} {\B{18}} \right) \cdot 
\left( 1-C \binom {e_{6}} {\B{18}} \right) \nonumber \\
& \left( 1+C \binom {z_1} {\B{18}} \right) \cdot 
\left( 1+C \binom {\alpha} {\B{18}} \right) \nonumber
\end{align*}
The corresponding matrix equations are given as:

\begin{align*}
\begin{pmatrix} (e_1|e_3)&(e_1|e_4)&(e_1|e_5)&(e_1|e_6)\\
(e_2|e_3)&(e_2|e_4)&(e_2|e_5)&(e_2|e_6)\\
(z_1|e_3)&(z_1|e_4)&(z_1|e_5)&(z_1|e_6)\\
(\alpha|e_3)&(\alpha|e_4)&(\alpha|e_5)&(\alpha|e_6)
\end{pmatrix}
\begin{pmatrix} p\\q\\r\\s\end{pmatrix} &=
\begin{pmatrix} (e_1|b_1 + z_1 + z_2 + \alpha)\\
(e_2|b_1 + z_1 + z_2 + \alpha)\\
(z_1|b_1 + z_1 + z_2 + \alpha) + 1\\
(\alpha|b_1 + z_1 + z_2 + \alpha) + 1\\
\end{pmatrix} \nonumber
\\[0.3cm]
\begin{pmatrix} (e_3|e_1)&(e_3|e_2)&(e_3|e_5)&(e_3|e_6)\\
(e_4|e_1)&(e_4|e_2)&(e_4|e_5)&(e_4|e_6)\\
(z_1|e_1)&(z_1|e_2)&(z_1|e_5)&(z_1|e_6)\\
(\alpha|e_1)&(\alpha|e_2)&(\alpha|e_5)&(\alpha|e_6)
\end{pmatrix}
\begin{pmatrix} p\\q\\r\\s\end{pmatrix} &=
\begin{pmatrix} (e_3|b_2 + z_1 + z_2 + \alpha)\\
(e_4|b_2 + z_1 + z_2 + \alpha)\\
(z_1|b_2 + z_1 + z_2 + \alpha) + 1\\
(\alpha|b_2 + z_1 + z_2 + \alpha) + 1\\
\end{pmatrix}
\\[0.3cm]
\begin{pmatrix} (e_5|e_1)&(e_5|e_2)&(e_5|e_3)&(e_5|e_4)\\
(e_6|e_1)&(e_6|e_2)&(e_6|e_3)&(e_6|e_4)\\
(z_1|e_1)&(z_1|e_2)&(z_1|e_3)&(z_1|e_4)\\
(\alpha|e_1)&(\alpha|e_2)&(\alpha|e_3)&(\alpha|e_4)
\end{pmatrix}
\begin{pmatrix} p\\q\\r\\s\end{pmatrix} &=
\begin{pmatrix} (e_5|b_1 + b_2 + z_2 + \alpha)\\
(e_6|b_1 + b_2 + z_2 + \alpha)\\
(z_1|b_1 + b_2 + z_2 + \alpha) + 1\\
(\alpha|b_1 + b_2 + z_2 + \alpha) + 1\\
\end{pmatrix} \nonumber
\end{align*}
The sectors 
\begin{align*}
\B{19,20,21}&=\B{1,2,3} + \alpha
\end{align*}
produce massless states that are obtained by acting 
on the vacuum with a fermionic oscillator. 
Below we list the type of states that are produced
and the projectors that act on them. 
\begin{itemize}

\item
\textbf{States $\{\overline\eta^{1}\}|R\rangle$ and $\{\overline\psi^{1,...,5}\}|R\rangle$}

These transform as either singlets or $\bf{5}$ or $\overline{\bf{5}}$ under the 
observable $SU(5)$ gauge group. The projectors are given by: 

\begin{align*}
	{P_{pqrs}^{(19)({\overline\eta^{1},\overline\psi}^{1,...,5})}} = \frac{1}{16} 
&\left( 1-C \binom {e_{1}} {\B{19}}\right) \cdot \left( 1-C \binom {e_{2}} 
{\B{19}}\right) \nonumber \\
	\cdot &\left( 1+C \binom {z_1} {\B{19}}\right) \cdot 
\left( 1 + C \binom {z_2 + \alpha}
	{\B{19}}\right) \nonumber \\
	{P_{pqrs}^{(20)({\overline\eta^{1},\overline\psi}^{1,...,5})}} = \frac{1}{16} 
&\left( 1-C \binom {e_{3}} {\B{20}}\right) \cdot \left( 1-C \binom {e_{4}} 
{\B{20}}\right) \nonumber \\
	\cdot &\left( 1+C \binom {z_1} {\B{20}}\right) \cdot 
\left( 1 + C \binom {z_2 + \alpha}
 	{\B{20}}\right) \\
  	{P_{pqrs}^{(21)({\overline\eta^{1},\overline\psi}^{1,...,5})}} = \frac{1}{16} 
&\left( 1-C \binom {e_{5}} {\B{21}}\right) \cdot \left( 1-C \binom {e_{6}} 
{\B{21}}\right) \nonumber \\
  	\cdot &\left( 1+C \binom {z_1} {\B{21}}\right) \cdot 
\left( 1 + C \binom {z_2 + \alpha}
   	{\B{21}}\right) \nonumber
\end{align*}
The corresponding matrix equations are given as:

\begin{align*}
\begin{pmatrix} (e_1|e_3)&(e_1|e_4)&(e_1|e_5)&(e_1|e_6)\\
(e_2|e_3)&(e_2|e_4)&(e_2|e_5)&(e_2|e_6)\\
(z_1|e_3)&(z_1|e_4)&(z_1|e_5)&(z_1|e_6)\\
(\delta|e_3)&(\delta|e_4)&(\delta|e_5)&(\delta|e_6)\\
\end{pmatrix}
\begin{pmatrix} p\\q\\r\\s\end{pmatrix} &=
\begin{pmatrix} (e_1|b_1+\alpha)\\
(e_2|b_1+\alpha)\\
(z_1|b_1+\alpha) + 1\\
(z_2|b_1) + (\alpha|b_1+ z_2 + \alpha) + 1
\end{pmatrix} \nonumber
\\[0.3cm]
\begin{pmatrix} (e_3|e_1)&(e_3|e_2)&(e_3|e_5)&(e_3|e_6)\\
(e_4|e_1)&(e_4|e_2)&(e_4|e_5)&(e_4|e_6)\\
(z_1|e_1)&(z_1|e_2)&(z_1|e_5)&(z_1|e_6)\\
(\delta|e_1)&(\delta|e_2)&(\delta|e_5)&(\delta|e_6)\\
\end{pmatrix}
\begin{pmatrix} p\\q\\r\\s\end{pmatrix} &=
\begin{pmatrix} (e_3|b_2+\alpha)\\
(e_4|b_2+\alpha)\\
(z_1|b_2+\alpha) + 1\\
(z_2|b_2) + (\alpha|b_2+ z_2 + \alpha) + 1
\end{pmatrix}
\\[0.3cm]
\begin{pmatrix} (e_5|e_1)&(e_5|e_2)&(e_5|e_3)&(e_5|e_4)\\
(e_6|e_1)&(e_6|e_2)&(e_6|e_3)&(e_6|e_4)\\
(z_1|e_1)&(z_1|e_2)&(z_1|e_3)&(z_1|e_4)\\
(\delta|e_1)&(\delta|e_2)&(\delta|e_3)&(\delta|e_4)\\
\end{pmatrix}
\begin{pmatrix} p\\q\\r\\s\end{pmatrix} &=
\begin{pmatrix} (e_5|b_3+\alpha)\\
(e_6|b_3+\alpha)\\
(z_1|b_3+\alpha) + 1\\
(z_2|b_3) + (\alpha|b_3 + z_2 + \alpha) + 1
\end{pmatrix} \nonumber
\end{align*}
Where $\delta = z_2 + \alpha$

\newpage
\item
\textbf{States $\{\overline\eta^{*2,3}\}|R\rangle$}

These transform as singlets under the observable 
$SU(5)$ gauge group and are charged under $U(1)_{2/3}$. 
The projectors are given by:
\begin{align*}
	{P_{pqrs}^{(19)({\overline\eta^{*2,3}})}} = \frac{1}{16} 
&\left( 1-C \binom {e_{1}} {\B{19}}\right) \cdot 
\left( 1-C \binom {e_{2}} {\B{19}}\right) \nonumber \\
	\cdot 
&\left( 1+C \binom {z_1} {\B{19}}\right) \cdot \left( 1 - C \binom {z_2 + \alpha}
	{\B{19}}\right) \nonumber \\
	{P_{pqrs}^{(20)({\overline\eta^{*2,3}})}} = \frac{1}{16} 
&\left( 1-C \binom {e_{3}} {\B{20}}\right) \cdot \left( 1-C \binom {e_{4}} 
{\B{20}}\right) \nonumber \\
	\cdot &\left( 1+C \binom {z_1} {\B{20}}\right) \cdot 
\left( 1 - C \binom {z_2 + \alpha}
 	{\B{20}}\right) \\
  	{P_{pqrs}^{(21)({\overline\eta^{*2,3}})}} = \frac{1}{16} 
&\left( 1-C \binom {e_{5}} {\B{21}}\right) \cdot \left( 1-C \binom {e_{6}} 
{\B{21}}\right) \nonumber \\
  	\cdot &\left( 1+C \binom {z_1} {\B{21}}\right) \cdot 
\left( 1 - C \binom {z_2 + \alpha}
   	{\B{21}}\right) \nonumber
\end{align*}
The corresponding matrix equations are given as:

\begin{align*}
\begin{pmatrix} (e_1|e_3)&(e_1|e_4)&(e_1|e_5)&(e_1|e_6)\\
(e_2|e_3)&(e_2|e_4)&(e_2|e_5)&(e_2|e_6)\\
(z_1|e_3)&(z_1|e_4)&(z_1|e_5)&(z_1|e_6)\\
(\delta|e_3)&(\delta|e_4)&(\delta|e_5)&(\delta|e_6)\\
\end{pmatrix}
\begin{pmatrix} p\\q\\r\\s\end{pmatrix} &=
\begin{pmatrix} (e_1|b_1+\alpha)\\
(e_2|b_1+\alpha)\\
(z_1|b_1+\alpha) + 1\\
(z_2|b_1) + (\alpha|b_1+ z_2 + \alpha)
\end{pmatrix} \nonumber
\\[0.3cm]
\begin{pmatrix} (e_3|e_1)&(e_3|e_2)&(e_3|e_5)&(e_3|e_6)\\
(e_4|e_1)&(e_4|e_2)&(e_4|e_5)&(e_4|e_6)\\
(z_1|e_1)&(z_1|e_2)&(z_1|e_5)&(z_1|e_6)\\
(\delta|e_1)&(\delta|e_2)&(\delta|e_5)&(\delta|e_6)\\
\end{pmatrix}
\begin{pmatrix} p\\q\\r\\s\end{pmatrix} &=
\begin{pmatrix} (e_3|b_2+\alpha)\\
(e_4|b_2+\alpha)\\
(z_1|b_2+\alpha) + 1\\
(z_2|b_2) + (\alpha|b_2+ z_2 + \alpha)
\end{pmatrix}
\\[0.3cm]
\begin{pmatrix} (e_5|e_1)&(e_5|e_2)&(e_5|e_3)&(e_5|e_4)\\
(e_6|e_1)&(e_6|e_2)&(e_6|e_3)&(e_6|e_4)\\
(z_1|e_1)&(z_1|e_2)&(z_1|e_3)&(z_1|e_4)\\
(\delta|e_1)&(\delta|e_2)&(\delta|e_3)&(\delta|e_4)\\
\end{pmatrix}
\begin{pmatrix} p\\q\\r\\s\end{pmatrix} &=
\begin{pmatrix} (e_5|b_3+\alpha)\\
(e_6|b_3+\alpha)\\
(z_1|b_3+\alpha) + 1\\
(z_2|b_3) + (\alpha|b_3 + z_2 + \alpha)
\end{pmatrix} \nonumber
\end{align*}
Where $\delta = z_2 + \alpha$

\newpage
\item
\textbf{States $\{\overline\phi^{*1,...,4}\}|R\rangle$}

These transform as singlets under the observable
$SU(5)$ gauge group 
and transform in non--trivial representation of the hidden
$SU(4) \times U(1)_4$ gauge group. The projectors are given by: 
\begin{align*}
	{P_{pqrs}^{(19)({\overline\phi^{*1,...,4}})}} = \frac{1}{16} 
&\left( 1-C \binom {e_{1}} {\B{19}}\right) \cdot 
\left( 1-C \binom {e_{2}} {\B{19}}\right) \nonumber \\
	\cdot &\left( 1 - C \binom {z_1} 
{\B{19}}\right) \cdot \left( 1 - C \binom {z_2 + \alpha}
	{\B{19}}\right) \nonumber \\
	{P_{pqrs}^{(20)({\overline\phi^{*1,...,4}})}} = \frac{1}{16} 
&\left( 1-C \binom {e_{3}} {\B{20}}\right) \cdot \left( 1-C \binom {e_{4}} 
{\B{20}}\right) \nonumber \\
	\cdot &\left( 1 - C \binom {z_1} {\B{20}}\right) \cdot 
\left( 1 - C \binom {z_2 + \alpha}
 	{\B{20}}\right) \\
  	{P_{pqrs}^{(21)({\overline\phi^{*1,...,4}})}} = \frac{1}{16} 
&\left( 1-C \binom {e_{5}} {\B{21}}\right) \cdot
 \left( 1-C \binom {e_{6}} {\B{21}}\right) \nonumber \\
  	\cdot &\left( 1 - C \binom {z_1} {\B{21}}\right) \cdot 
\left( 1 - C \binom {z_2 + \alpha}
   	{\B{21}}\right) \nonumber
\end{align*}
The corresponding matrix equations are given as:

\begin{align*}
\begin{pmatrix} (e_1|e_3)&(e_1|e_4)&(e_1|e_5)&(e_1|e_6)\\
(e_2|e_3)&(e_2|e_4)&(e_2|e_5)&(e_2|e_6)\\
(z_1|e_3)&(z_1|e_4)&(z_1|e_5)&(z_1|e_6)\\
(\delta|e_3)&(\delta|e_4)&(\delta|e_5)&(\delta|e_6)\\
\end{pmatrix}
\begin{pmatrix} p\\q\\r\\s\end{pmatrix} &=
\begin{pmatrix} (e_1|b_1+\alpha)\\
(e_2|b_1+\alpha)\\
(z_1|b_1+\alpha) \\
(z_2|b_1) + (\alpha|b_1+ z_2 + \alpha) \\
\end{pmatrix} \nonumber
\\[0.3cm]
\begin{pmatrix} (e_3|e_1)&(e_3|e_2)&(e_3|e_5)&(e_3|e_6)\\
(e_4|e_1)&(e_4|e_2)&(e_4|e_5)&(e_4|e_6)\\
(z_1|e_1)&(z_1|e_2)&(z_1|e_5)&(z_1|e_6)\\
(\delta|e_1)&(\delta|e_2)&(\delta|e_5)&(\delta|e_6)\\
\end{pmatrix}
\begin{pmatrix} p\\q\\r\\s\end{pmatrix} &=
\begin{pmatrix} (e_3|b_2+\alpha)\\
(e_4|b_2+\alpha)\\
(z_1|b_2+\alpha) \\
(z_2|b_2) + (\alpha|b_2+ z_2 + \alpha) \\
\end{pmatrix}
\\[0.3cm]
\begin{pmatrix} (e_5|e_1)&(e_5|e_2)&(e_5|e_3)&(e_5|e_4)\\
(e_6|e_1)&(e_6|e_2)&(e_6|e_3)&(e_6|e_4)\\
(z_1|e_1)&(z_1|e_2)&(z_1|e_3)&(z_1|e_4)\\
(\delta|e_1)&(\delta|e_2)&(\delta|e_3)&(\delta|e_4)\\
\end{pmatrix}
\begin{pmatrix} p\\q\\r\\s\end{pmatrix} &=
\begin{pmatrix} (e_5|b_3+\alpha)\\
(e_6|b_3+\alpha)\\
(z_1|b_3+\alpha) \\
(z_2|b_3) + (\alpha|b_3+ z_2 + \alpha)
\end{pmatrix} \nonumber
\end{align*}
Where $\delta = z_2 + \alpha$

\end{itemize}
The remaining sectors
\begin{align*}
\B{22,23,24}&=\B{1,2,3} + z_1 + \alpha
\end{align*}
produce the following vector-like states:

\begin{itemize}
\item
\textbf{States $\{\overline\eta^{1}\}|R\rangle$ and $\{\overline\psi^{1,...,5}\}|R\rangle$}

These transform as either singlets or $\bf{5}$ or $\overline{\bf{5}}$ under the observable
$SU(5)$ gauge group. 
The projectors are given by:

\begin{align*}
{P_{pqrs}^{(22)({\overline\eta^{1},\overline\psi}^{1,...,5})}} = \frac{1}{16} 
&\left( 1-C \binom {e_{1}} {\B{22}}\right) \cdot \left( 1-C \binom {e_{2}} 
{\B{22}}\right) \,\,\,\, \nonumber \\
\cdot &\left( 1+C \binom {z_1} {\B{22}}\right) \cdot 
\left( 1 - C \binom {z_2 + \alpha} {\B{22}} \right)  \nonumber \\
{P_{pqrs}^{(23)({\overline\eta^{1},\overline\psi}^{1,...,5})}} = \frac{1}{16} 
&\left( 1-C \binom {e_{3}} {\B{23}}\right) \cdot \left( 1-C \binom {e_{4}} 
{\B{23}}\right) \,\,\,\, \nonumber \\
\cdot &\left( 1+C \binom {z_1} {\B{23}}\right) \cdot 
\left( 1 - C \binom {z_2 + \alpha} {\B{23}} \right) \\
{P_{pqrs}^{(24)({\overline\eta^{1},\overline\psi}^{1,...,5})}} = \frac{1}{16} 
&\left( 1-C \binom {e_{5}} {\B{24}}\right) \cdot 
\left( 1-C \binom {e_{6}} {\B{24}}\right) \,\,\,\, \nonumber \\
\cdot &\left( 1+C \binom {z_1} {\B{24}}\right) \cdot 
\left( 1 - C \binom {z_2 + \alpha} {\B{24}} \right)  \nonumber
\end{align*}
The corresponding matrix equations are given as:

\begin{align*}
\begin{pmatrix} (e_1|e_3)&(e_1|e_4)&(e_1|e_5)&(e_1|e_6)\\
(e_2|e_3)&(e_2|e_4)&(e_2|e_5)&(e_2|e_6)\\
(z_1|e_3)&(z_1|e_4)&(z_1|e_5)&(z_1|e_6)\\
({\delta}|e_3)&({\delta}|e_4)&({\delta}|e_5)&({\delta}|e_6)\\
\end{pmatrix}
\begin{pmatrix} p \\ q \\ r \\ s \end{pmatrix} &=
\begin{pmatrix} (e_1|b_1 + z_1 + \alpha)\\
(e_2|b_1 + z_1 + \alpha)\\
(z_1|b_1 + z_1 + \alpha) + 1\\
({z_2}|b_1 + z_1) + (\alpha|b_1 + z_1 + z_2 + \alpha)\\
\end{pmatrix} \nonumber \\[0.3cm]
\begin{pmatrix} (e_3|e_1)&(e_3|e_2)&(e_3|e_5)&(e_3|e_6)\\
(e_4|e_1)&(e_4|e_2)&(e_4|e_5)&(e_4|e_6)\\
(z_1|e_1)&(z_1|e_2)&(z_1|e_5)&(z_1|e_6)\\
({\delta}|e_1)&({\delta}|e_2)&({\delta}|e_5)&({\delta}|e_6)\\
\end{pmatrix}
\begin{pmatrix} p \\ q \\ r \\ s \end{pmatrix} &=
\begin{pmatrix} (e_3|b_2 + z_1 + \alpha)\\
(e_4|b_2 + z_1 + \alpha)\\
(z_1|b_2 + z_1 + \alpha) + 1\\
({z_2}|b_2 + z_1) + (\alpha|b_2 + z_1 + z_2 + \alpha)\\
\end{pmatrix}  \\[0.3cm] 
\begin{pmatrix} (e_5|e_1)&(e_5|e_2)&(e_5|e_3)&(e_5|e_4)\\
(e_6|e_1)&(e_6|e_2)&(e_6|e_3)&(e_6|e_4)\\
(z_1|e_1)&(z_1|e_2)&(z_1|e_3)&(z_1|e_4)\\
({\delta}|e_1)&({\delta}|e_2)&({\delta}|e_3)&({\delta}|e_4)\\
\end{pmatrix}
\begin{pmatrix} p \\ q \\ r \\ s \end{pmatrix} &=
\begin{pmatrix} (e_5|b_3 + z_1 + \alpha)\\
(e_6|b_3 + z_1 + \alpha)\\
(z_1|b_3 + z_1 + \alpha) + 1\\
({z_2}|b_1 + b_2) + (\alpha|b_1 + b_2 + z_2 + \alpha) + 1
\end{pmatrix} \nonumber
\end{align*}
Where $\delta = z_2 + \alpha$
\newpage
\item
\textbf{States $\{\overline\eta^{*2,3}\}|R\rangle$}

These transform as singlets under the observable
$SU(5)$ gauge group and are charged under $U(1)_{2/3}$. 
The projectors are given by: 

\begin{align*}
{P_{pqrs}^{(22)({\overline\eta^{*2,3}})}} = \frac{1}{16} 
&\left( 1-C \binom {e_{1}} {\B{22}}\right) \cdot 
\left( 1-C \binom {e_{2}} {\B{22}}\right) \,\,\,\, \nonumber \\
\cdot &\left( 1+C \binom {z_1} {\B{22}}\right) \cdot 
\left( 1 + C \binom {z_2 + \alpha} {\B{22}}\right)  \nonumber \\
{P_{pqrs}^{(23)({\overline\eta^{*2,3}})}} = \frac{1}{16} 
&\left( 1-C \binom {e_{3}} {\B{23}}\right) \cdot 
\left( 1-C \binom {e_{4}} {\B{23}}\right) \,\,\,\, \nonumber \\
\cdot &\left( 1+C \binom {z_1} {\B{23}}\right) \cdot 
\left( 1 + C \binom {z_2 + \alpha} {\B{23}}\right) \\
{P_{pqrs}^{(24)({\overline\eta^{*2,3}})}} = \frac{1}{16} 
&\left( 1-C \binom {e_{5}} {\B{24}}\right) \cdot 
\left( 1-C \binom {e_{6}} {\B{24}}\right) \,\,\,\, \nonumber \\
\cdot &\left( 1+C \binom {z_1} {\B{24}}\right) \cdot 
\left( 1 + C \binom {z_2 + \alpha} {\B{24}}\right)  \nonumber
\end{align*}
The corresponding matrix equations are given as:

\begin{align*}
\begin{pmatrix} (e_1|e_3)&(e_1|e_4)&(e_1|e_5)&(e_1|e_6)\\
(e_2|e_3)&(e_2|e_4)&(e_2|e_5)&(e_2|e_6)\\
(z_1|e_3)&(z_1|e_4)&(z_1|e_5)&(z_1|e_6)\\
({\delta}|e_3)&({\delta}|e_4)&({\delta}|e_5)&({\delta}|e_6)\\
\end{pmatrix}
\begin{pmatrix} p \\ q \\ r \\ s \end{pmatrix} &=
\begin{pmatrix} (e_1|b_1 + z_1 + \alpha)\\
(e_2|b_1 + z_1 + \alpha)\\
(z_1|b_1 + z_1 + \alpha) + 1\\
({z_2}|b_1 + z_1) + (\alpha|b_1 + z_1 + z_2 + \alpha) + 1\\
\end{pmatrix} \nonumber \\[0.3cm]
\begin{pmatrix} (e_3|e_1)&(e_3|e_2)&(e_3|e_5)&(e_3|e_6)\\
(e_4|e_1)&(e_4|e_2)&(e_4|e_5)&(e_4|e_6)\\
(z_1|e_1)&(z_1|e_2)&(z_1|e_5)&(z_1|e_6)\\
({\delta}|e_1)&({\delta}|e_2)&({\delta}|e_5)&({\delta}|e_6)\\
\end{pmatrix}
\begin{pmatrix} p \\ q \\ r \\ s \end{pmatrix} &=
\begin{pmatrix} (e_3|b_2 + z_1 + \alpha)\\
(e_4|b_2 + z_1 + \alpha)\\
(z_1|b_2 + z_1 + \alpha) + 1\\
({z_2}|b_2 + z_1) + (\alpha|b_2 + z_1 + z_2 + \alpha) + 1\\
\end{pmatrix}  \\[0.3cm] 
\begin{pmatrix} (e_5|e_1)&(e_5|e_2)&(e_5|e_3)&(e_5|e_4)\\
(e_6|e_1)&(e_6|e_2)&(e_6|e_3)&(e_6|e_4)\\
(z_1|e_1)&(z_1|e_2)&(z_1|e_3)&(z_1|e_4)\\
({\delta}|e_1)&({\delta}|e_2)&({\delta}|e_3)&({\delta}|e_4)\\
\end{pmatrix}
\begin{pmatrix} p \\ q \\ r \\ s \end{pmatrix} &=
\begin{pmatrix} (e_5|b_3 + z_1 + \alpha)\\
(e_6|b_3 + z_1 + \alpha)\\
(z_1|b_3 + z_1 + \alpha) + 1\\
({z_2}|b_1 + b_2) + (\alpha|b_1 + b_2 + z_2 + \alpha)
\end{pmatrix} \nonumber
\end{align*}
Where $\delta = z_2 + \alpha$

\newpage
\item
\textbf{States $\{\overline\phi^{1,...,4}\}|R\rangle$}

These transform as singlets under the observable $SU(5)$ gauge group 
and transform in non--trivial representations of the 
the hidden $SU(4) \times U(1)_4$ gauge group.
The projectors are given by: 

\begin{align*}
{P_{pqrs}^{(22)({\overline\phi^{1,...,4}})}} = \frac{1}{16} 
&\left( 1-C \binom {e_{1}} {\B{22}}\right) \cdot 
\left( 1-C \binom {e_{2}} {\B{22}}\right) \,\,\,\, \nonumber \\
\cdot &\left( 1 - C \binom {z_1} {\B{22}}\right) \cdot 
\left( 1+ C \binom {z_2 + \alpha} {\B{22}}\right)  \nonumber \\
{P_{pqrs}^{(23)({\overline\phi^{1,...,4}})}} = \frac{1}{16} 
&\left( 1-C \binom {e_{3}} {\B{23}}\right) \cdot 
\left( 1-C \binom {e_{4}} {\B{23}}\right) \,\,\,\, \nonumber \\
\cdot &\left( 1 - C \binom {z_1} {\B{23}}\right) \cdot 
\left( 1+ C \binom {z_2 + \alpha} {\B{23}}\right) \\
{P_{pqrs}^{(24)({\overline\phi^{1,...,4}})}} = \frac{1}{16} 
&\left( 1-C \binom {e_{5}} {\B{24}}\right) \cdot 
\left( 1-C \binom {e_{6}} {\B{24}}\right) \,\,\,\, \nonumber \\
\cdot &\left( 1 - C \binom {z_1} {\B{24}}\right) 
\cdot \left( 1+ C \binom {z_2 + \alpha} {\B{24}}\right)  \nonumber
\end{align*}
The corresponding matrix equations are given as:

\begin{align*}
\begin{pmatrix} (e_1|e_3)&(e_1|e_4)&(e_1|e_5)&(e_1|e_6)\\
(e_2|e_3)&(e_2|e_4)&(e_2|e_5)&(e_2|e_6)\\
(z_1|e_3)&(z_1|e_4)&(z_1|e_5)&(z_1|e_6)\\
({\delta}|e_3)&({\delta}|e_4)&({\delta}|e_5)&({\delta}|e_6)\\
\end{pmatrix}
\begin{pmatrix} p \\ q \\ r \\ s \end{pmatrix} &=
\begin{pmatrix} (e_1|b_1 + z_1 + \alpha)\\
(e_2|b_1 + z_1 + \alpha)\\
(z_1|b_1 + z_1 + \alpha)\\
({z_2}|b_1 + z_1) + (\alpha|b_1 + z_1 + z_2 + \alpha)\\
\end{pmatrix} \nonumber \\[0.3cm]
\begin{pmatrix} (e_3|e_1)&(e_3|e_2)&(e_3|e_5)&(e_3|e_6)\\
(e_4|e_1)&(e_4|e_2)&(e_4|e_5)&(e_4|e_6)\\
(z_1|e_1)&(z_1|e_2)&(z_1|e_5)&(z_1|e_6)\\
({\delta}|e_1)&({\delta}|e_2)&({\delta}|e_5)&({\delta}|e_6)\\
\end{pmatrix}
\begin{pmatrix} p \\ q \\ r \\ s \end{pmatrix} &=
\begin{pmatrix} (e_3|b_2 + z_1 + \alpha)\\
(e_4|b_2 + z_1 + \alpha)\\
(z_1|b_2 + z_1 + \alpha)\\
({z_2}|b_2 + z_1) + (\alpha|b_2 + z_1 + z_2 + \alpha)\\
\end{pmatrix}  \\[0.3cm] 
\begin{pmatrix} (e_5|e_1)&(e_5|e_2)&(e_5|e_3)&(e_5|e_4)\\
(e_6|e_1)&(e_6|e_2)&(e_6|e_3)&(e_6|e_4)\\
(z_1|e_1)&(z_1|e_2)&(z_1|e_3)&(z_1|e_4)\\
({\delta}|e_1)&({\delta}|e_2)&({\delta}|e_3)&({\delta}|e_4)\\
\end{pmatrix}
\begin{pmatrix} p \\ q \\ r \\ s \end{pmatrix} &=
\begin{pmatrix} (e_5|b_3 + z_1 + \alpha)\\
(e_6|b_3 + z_1 + \alpha)\\
(z_1|b_3 + z_1 + \alpha)\\
({z_2}|b_1 + b_2) + (\alpha|b_1 + b_2 + z_2 + \alpha) + 1
\end{pmatrix} \nonumber
\end{align*}
Where $\delta = z_2 + \alpha$

\end{itemize}

\newpage

\bigskip
\medskip

\bibliographystyle{unsrt}

\end{document}